\documentclass[pdflatex,sn-mathphys-num]{sn-jnl}

\usepackage{bm}
\usepackage{bbm}

\usepackage{anyfontsize}

\usepackage{graphicx}%
\usepackage{multirow}%
\usepackage{amsmath,amssymb,amsfonts}%
\usepackage{amsthm}%
\usepackage{mathrsfs}%
\usepackage[title]{appendix}%
\usepackage{xcolor}%
\usepackage{textcomp}%
\usepackage{manyfoot}%
\usepackage{booktabs}%
\usepackage{algorithm}%
\usepackage{algorithmicx}%
\usepackage{algpseudocode}%
\usepackage{listings}%
\usepackage{pgfplots}
\pgfplotsset{compat=newest}
\usepgfplotslibrary{groupplots}
\usepgfplotslibrary{dateplot}

\newtheorem{remark}{Remark}

\begin{document}

\title[Article Title]{Statistical reduced order modelling for the parametric Helmholtz equation}


\author*[1]{\fnm{Lucas} \sur{Hermann}}\email{l.hermann@tu-braunschweig.de}

\author[2]{\fnm{Matthias} \sur{Bollhöfer}}\email{m.bollhoefer@tu-braunschweig.de}

\author[1]{\fnm{Ulrich} \sur{Römer}}\email{u.roemer@tu-braunschweig.de}

\affil*[1]{\orgdiv{Institute for Acoustics and Dynamics}, \orgname{Technische Universität Braunschweig}, \orgaddress{\street{Langer Kamp 19}, \city{Braunschweig}, \postcode{38106}, \state{NI}, \country{Germany}}}

\affil[2]{\orgdiv{Institute for Numerical Analysis}, \orgname{Technische Universität Braunschweig}, \orgaddress{\street{Universitätsplatz 2}, \city{Braunschweig}, \postcode{38106}, \state{NI}, \country{Germany}}}


\abstract{Predictive modelling combining both numerical simulations and real-world measurement data, obtained from sensors, is gaining importance in computational science and engineering. Even with large-scale finite element models, a mismatch to the sensor data often remains, which can be attributed to different sources of uncertainty. For such a scenario, the statistical finite element method (statFEM) can be used to condition a simulated field on given sensor data. This yields a posterior solution which resembles the data much better and additionally provides consistent estimates of uncertainty, including model misspecification. For frequency or parameter dependent problems, occurring, e.g. in acoustics or electromagnetism, solving the full order model across the frequency range of interest and conditioning it on data quickly results in a prohibitive computational cost. In this case, the introduction of a surrogate in the form of a reduced order model (ROM) yields much smaller systems of equations. In this paper, we propose a reduced order statFEM framework relying on Krylov-based moment matching. We introduce a data model which explicitly includes the bias induced by the reduced approximation, which is estimated by an 
error indicator. The results of the new statistical reduced order method are compared to the standard statFEM procedure applied to a ROM prior without explicitly accounting for reduced order bias. The proposed method achieves better accuracy and faster convergence throughout a given frequency range for a variety of numerical examples and artificial (i.e. simulated) sensor data. }

\keywords{statFEM, moment matching, Gaussian Processes, adjoint error estimator}



\maketitle

\section{Introduction}\label{sec1}

In engineering and physics, but also in many other disciplines such as chemistry and finance, partial differential equations (PDEs) \cite{Borthwick2016} serve as a very fundamental mathematical description of a given problem. Generally, most PDEs are hard or impossible to solve analytically. To find an approximation of the solution, numerical methods such as the finite element method (FEM) \cite{Brenner2008} are used. Both the model assumptions, i.e. choice of the differential operator, parameters and boundary conditions, which lead to the PDE and the approximative nature of the numerical method contribute to the difference between the solution of the numerical model and physical reality. Collecting data with sensors may yield relatively accurate, albeit noisy, observations at discrete locations in the computational domain. The numerical solution, on the other hand, yields a continuous solution at every point in the domain. Combining the numerical solution with the sparse observation data to estimate the state of a system is referred to as \emph{data assimilation} \cite{law2015}. 

One data assimilation technique is the recently proposed statistical FEM (statFEM) \cite{girolami2021statistical}. In statFEM, a statistical model for measured sensor data is introduced in which the data are additively composed of the numerical FE solution, a model discrepancy and measurement noise. All these terms are modeled as Gaussian Processes (GP) \cite{rasmussen2006}, which enables data assimilation through Bayesian principles. By propagating uncertainty through the FE model, a Gaussian prior for the state is obtained.
In the original statFEM paper, a first-order perturbation approach \cite{Liu1986-al} is employed, but Monte Carlo type methods are also well-suited. Once the prior is at hand, conditioning on measurement data leads to the posterior.  Despite acknowledging the uncertainty, the model may still be misspecified, in particular for complex physical processes. Hence, a model error GP is introduced, which is assumed to be mean-free, i.e. a possible discrepancy between model and reality is reflected in the variance of the model error. Adopting an empirical Bayes approach, the hyperparameters of the covariance function are learned from the data first, before the solution GP is conditioned on the data, where measurement error is also accounted for. 

Some engineering problems require solving PDEs depending on deterministic parameters in addition to the uncertain model inputs. This concerns frequency domain problems, i.e. the Helmholtz equation in acoustics \cite{ihlenburg1998}, Maxwell's equations in electromagnetism \cite{maxwell1865} or the equations of motion in structural dynamics \cite{Timoshenko1990}. Concerning the latter, possible use cases can be found e.g. in the domain of structural health monitoring, where the state of a large scale structure \cite{rytter1993, farrar2001} or smaller scale aerospace turbine components is constantly observed by means of sparsely located sensors and a structural dynamics model has to be updated for the full frequency range once new measurements come in \cite{Heffernan2022}.  The resulting FE models can become very complex and expensive to solve even for a single frequency. Hence, computing a parameter sweep, i.e. solving the system at many frequencies within a desired interval, quickly becomes prohibitively expensive. Through reduced order models (ROM), the solution of e.g. a FEM system of equations with a large number of degrees of freedom (DOF) can be approximated by solving a significantly smaller ROM. The most prominent techniques for model order reduction are projection based methods \cite{Benner2015}. Examples of those are proper orthogonal decomposition (POD) or Krylov subspace methods. As, in general, neither of these ROM methods are a perfectly accurate surrogate for the FEM, also the ROM approximation error has to be taken into account, see for instance \cite{silva2023reduced}. In contrast to the model-to-reality error, this ROM approximation error has a well-known structure, governed by the error, or residual, equation.

In this paper, a statistical reduced order modelling approach, referred to as \mbox{statROM}, is introduced. We show that this approach leads to a better approximation of the ground truth and faster convergence of the error. We obtain the prior by quasi Monte Carlo analysis \cite{MOROKOFF1995} based on a frequency-dependent ROM and assimilate sparse sensor data with Bayesian updating. The ROM is constructed with a moment matching Krylov subspace method. Of particular importance in this work is the ROM error and an adjoint based error estimator \cite{HARTMANN2008, Rouse2022} is derived which does not involve solving the full order system of equations. This additional error is then included into the data model. In principle, the ROM error could also be estimated as part of the model error. Here, this will be referred to as the standard approach. However, as it is possible to estimate the ROM error without solving the full system, it seems sensible to separate it from the general model error and hence have a better prior for the total error. 

\begin{figure}[ht]
\centering
\includegraphics[width=0.8\textwidth]{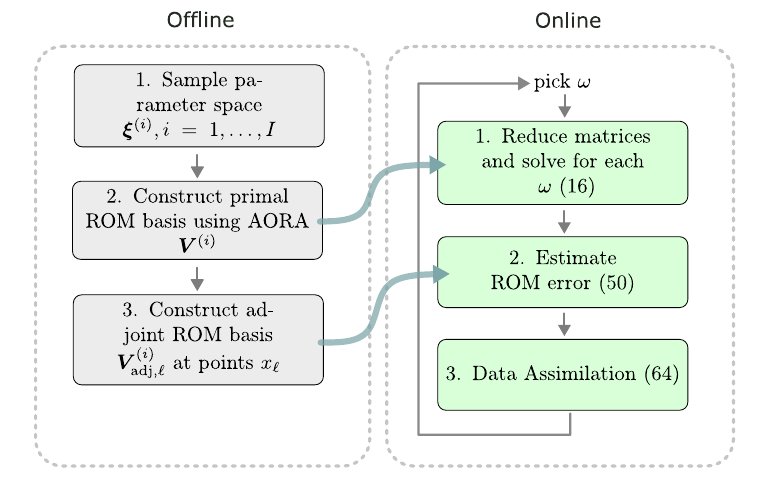}
\caption{Workflow diagram of the proposed statROM method. In an offline phase, the reduced order projection basis is constructed for the primal and adjoint problem and each sample of the input parameters. In an online phase the FE matrices are reduced and the system is solved in the reduced space at reduced cost compared to the full order model. The adjoint error indicator contributes to the quality of the subsequent data assimilation. Fast parameter sweeps over the frequency $\omega$ can be carried out in the online phase based on the sampled ROMs.}\label{fig:fig1}
\end{figure}
The method is applied to a FEM approximation of the Helmholtz equation in 1D with few DOFs to study the convergence properties in much detail. Data are generated from a fine FE discretisation. Next, a 2D Helmholtz scattering problem with significantly more DOFs is evaluated to study how the method scales. 

In Figure \ref{fig:fig1}, the workflow of the proposed method is sketched. A reduced order model that can be rapidly solved, enables an online phase with fast frequency sweeps. Our main contributions can be summarized as follows:
\begin{itemize}
    \item Extension of the statFEM method to handle parameter dependent problems with ROMs
    \item Introduction of a ROM error estimator to improve the posterior accuracy
    \item Application of statFEM and the proposed statROM method to the Helmholtz equation
\end{itemize}

The paper is structured as follows: In Section 2, the Helmholtz equation is introduced as the model problem. Section 3 outlines ROM and moment matching methods in particular. Also, an adjoint based ROM error estimator is derived. In Section 4, GPs and statFEM are introduced before deriving the statROM procedure. Section 5 is dedicated to numerical examples. Lastly, we draw a conclusion.

\section{Model problem}\label{sec2}
We consider the Helmholtz equation as an example of a frequency dependent PDE, which serves as a model problem, e.g. for wave-based non-destructive testing applications. In \cite{Willberg2015}, an overview over such methods is provided. Specific examples in the context of structural health monitoring using high fidelity FE models can be found in e.g. \cite{Rahmatalla2014,Giagopoulos2018,Mirasoli2022}. More specifically in the case treated here, the model PDE is given as 
\begin{align}
\label{eq:model_strong}
    \Delta u(\bm{x}) + k^2 \kappa(\bm{x}) u(\bm{x}) &= -f(\bm{x}), \quad &&\text{in } D, \\
    u(\bm{x}) &= 0, \quad &&\text{on } \Gamma_{\mathrm{D}}, \\
    \frac{\partial u(\bm{x}) }{\partial n} &= g(\bm{x}) , \quad &&\text{on } \Gamma_{\mathrm{N}}, \\
    \frac{\partial u(\bm{x}) }{\partial n} - \mathrm{i} k \beta u(\bm{x})  &= 0, \quad &&\text{on } \Gamma_{\mathrm{I}},
\end{align}
on a bounded domain $D \subset \mathbb{R}^d$ with $d=1,2,3$, where $\partial D$ is a Lipschitz boundary for $d \geq 2$ that can be well-approximated by polygons or polyhedra. Moreover, $\bm{n}$ denotes the outward normal to $D$. The PDE is solved for the sound pressure field $u(\bm{x})$. A sound soft boundary condition can be realized with the homogeneous Dirichlet boundary condition. The Neumann boundary can be separated into a homogeneous, i.e. sound hard reflecting part $\Gamma_{\mathrm{N},0}$ and an inhomogeneous part $\Gamma_{\mathrm{N},g}$, where $g\neq 0$. Damping can be considered by introducing the Robin boundary condition on $\Gamma_{\mathrm{I}}$, with the specific normalized acoustic admittance $\beta \in \mathbb{R}$. Note that $\Gamma_{\mathrm{D}}$ and/or $\Gamma_{\mathrm{I}}$ may be empty. Furthermore the wave number is defined as $\mathbb{R} \ni k = \frac{\omega}{c}$, with $\omega$ denoting the angular frequency and $c$ the speed of sound and $i$ is the imaginary unit. Several quantities in the model problem are assumed to be unknown and hence, exhibit uncertainty. These are the material parameter $\kappa$, the inhomogeneity in the Neumann boundary condition $g$ and also the forcing $f$.

The corresponding weak formulation is posed on the (complex) function space 
\begin{equation}
V = H^1(D) = \{ v \in L^2(D) \ | \ \|v \|_V^2 = \int_D \nabla v \cdot \nabla v^* \mathrm{d}x + \int_D v v^* \mathrm{d}x < \infty \},
\end{equation}
where $v^*$ denotes the complex conjugate of $v$, and the subspace 
\begin{equation}
V_0 =  \{ v \in V \ | \ v|_{\Gamma_{\mathrm{D}}} =0 \}.
\end{equation}
The weak formulation reads: find $u \in V_0$ such that 
\begin{equation}
s(u,v) - k^2 m(u,v) - i k \beta d(u,v) = F(v), \ \forall v \in V_0,
\end{equation}
where the different quantities are defined as 
\begin{align}
    s(u,v) &= \int_{D} \nabla u \cdot \nabla v^* \mathrm{d}x,  && \text{(stiffness)},\\
    m(u,v) &= \int_{D}  \kappa  u v^* \mathrm{d}x,  && \text{(mass)},\\
    d(u,v) &= \int_{\Gamma_\mathrm{I}} u v^* \mathrm{d}x,  && \text{(damping)},\\
    F(v) &= \int_{\Gamma_{\mathrm{N},g}} g v^* \mathrm{d}x + \int_{D} f v^* \mathrm{d}x&& \text{(forcing)},
\end{align}
see \cite{Atalla2015,IHLENBURG1995} for details on the weak formulation. 

We introduce a conforming tetrahedral discretization of the domain and lowest order, piece-wise continuous, basis functions. For details, the reader is referred to \cite{Brenner2008,LoggMardalEtAl2012}. Let $V_{0,h},V_h$ denote the corresponding discrete function spaces with and without homogeneous Dirichlet boundary condition. Then, the discrete formulation reads: find $u_h \in V_{0,h}$ such that 
\begin{equation}
s(u_h,v_h) - k^2 m(u_h,v_h) - i k \beta d(u_h,v_h) = F(v_h), \ \forall v_h \in V_{0,h}.
\label{eqn:2ndorderFem}
\end{equation}
To prove that for all non-resonant $\omega \in [\omega_{\text{min}},\omega_{\text{max}}]$ the weak formulation has a unique solution, continuity of $s,m,d,F$ and a G\r{a}rding inequality to invoke the Fredholm alternative are required. Details on the proof and the required assumptions can be found in \cite{ihlenburg1998} and will not be discussed in any detail here. 

After FE-discretization, we obtain the following system of equations 
\begin{equation}
\underbrace{(\bm{S} - k^2 \bm{M} - i k \beta \bm{D})}_{=:\bm{A}} \bm{u} = \bm{F},
\label{eqn:femSys}
\end{equation}
with $\bm{M}$ the mass matrix, $\bm{D}$ the damping matrix and $\bm{S}$ the stiffness matrix. In \eqref{eqn:femSys}, $\bm{A} \in \mathbb{C}^{N\times N}$ denotes the system matrix, $\bm{u} \in \mathbb{C}^{N}$ the nodal coefficients and $\bm{F} \in \mathbb{C}^{N}$ the source vector. 

At this stage, we also introduce random influences, which model unknown material parameters and forcings. More precisely, $\kappa,f,g$ will be modeled as random fields on a common probability space $(\Theta,\Sigma,\mathcal{P})$ with a sample space $\Theta$, a sigma algebra $\Sigma$ and probability measure $\mathcal{P}$. We obtain
\begin{equation}
 \bm{A}(\omega,\theta) \bm{u}(\omega,\theta) = \bm{F}(\omega,\theta), \ \forall \omega \in [\omega_{\text{min}},\omega_{\text{max}}] \subset \mathbb{R}_+
\end{equation}
almost surely, where $\theta \in \Theta$ denotes an elementary random outcome. The dependence of the Helmholtz equation on a high-dimensional parameter vector, originating from a random field approximation, poses additional challenges, which have recently attracted interest in the context of forward uncertainty quantification \cite{Spence2022}.



\section{Reduced Order Modelling}\label{sec3}
In many situations in practice, the size of the discrete system $N$ is very large, which makes it impossible to solve problem \eqref{eqn:femSys} for many different values of the angular frequency $\omega$ or to efficiently perform an uncertainty analysis. Therefore, from the FEM system of equations, a ROM is typically derived, which can be used for multi-query tasks. As the model problem is defined in frequency domain, an uncertainty analysis is especially expensive, because it needs to be repeated for each point in the frequency domain. With a suitable ROM that describes the underlying FE-problem accurately, a reduction of numerical cost can be achieved, which in turn enables time-critical applications where, for instance, the model has to be conditioned on new incoming sensor data very frequently.


Throughout the paper, projection based ROMs \cite{Benner2015} will be used, where a projection matrix $\bm{V}$ projects the system components onto a much smaller space through Galerkin projection. 
We will consider multiple ROMs, which are constructed for specific random realizations. Hence, we fix $\theta$ in this subsection and omit it, for simplicity. The FEM and corresponding ROM system read
\begin{align}\label{eqn:FEMROM}
\text{FEM}&: \quad        \bm{A}(\omega) \bm{u}(\omega) = \bm{F}(\omega), &&\\
\label{eqn:FEMROM2}  \text{ROM}&:  \quad     \bm{A}_r(\omega) \bm{u}_r(\omega) = \bm{F}_r(\omega), &&
\end{align}
where $\bm{A}_r(\omega) \in \mathbb{C}^{r\times r}, \bm{u}_r(\omega), \bm{F}_r(\omega) \in \mathbb{C}^{r}$ and $r\ll N$. The FE solution and the ROM solution vector are connected by a projection matrix $\bm{V}\in \mathbb{C}^{N\times r}$, as
\begin{equation}\label{eqn:projSol}
\bm{u}(\omega) \approx \bm{V}\bm{u}_r(\omega)
\end{equation}
as well as $\bm{A}_r = \bm{V}^* \bm{A} \bm{V}$ and $\bm{F}_r = \bm{V}^* \bm{F}$. Without loss of generality, the projection matrix satisfies $\bm{V}^* \bm{V} = \bm{I}$, hence it is orthogonal.



There are various methods available to construct the projection matrix. Proper Orthogonal Decomposition (POD) is based on so-called snapshots of the FE model. For these, the FE system is solved at a set of parameter samples and the resulting vectors are stored as columns in a matrix. The columns of $\bm{V}$ are then spanned by the first $r$ principal components. POD is not the most efficient method for the model problem under consideration. Hence, a different method, introduced in the next subsection, will be used in this work.



A class of projection based ROMs which is well suited for dynamical systems \cite{Roemer2021} consists of moment matching methods. Moment matching is based on rational interpolation \cite{berrut2005} and the resulting ROM satisfies the Hermite interpolation conditions up to some order \cite{Benner2015}. The interpolatory approach can be considered in a structure preserving way within a much more general setting, including the second order dynamical system case as shown in \cite{BeaGu09}.
In particular, since \eqref{eqn:femSys} is a second order dynamical system, a structure-preserving second order moment matching method analogous to the SOAR procedure \cite{BaiSu05SC} will be used. To do so, we will use second order generalized Krylov subspaces \cite{BaiSu05,BaiSu05SC} applied to \eqref{eqn:femSys}. In the case of a single expansion frequency $\bar{\omega}$ we have that
\begin{equation}\label{eqn:soks}
    \mathcal{K}(\bm{M},\bm{D},\bm{S},\bm{F}) = G_m(-\bm{A}(\bar{\omega})^{-1}\bm{A}'(\bar{\omega}),-\bm{A}(\bar{\omega})^{-1} i \bm{M},\bm{A}(\bar{\omega})^{-1} \bm{F}(\bar{\omega}) )
\end{equation}
with 

\begin{multline}
    G_m(\bm{L},\bm{B},\bm{r}) = \mathrm{span}\{\bm{r}_0,\bm{r}_1,\dots,\bm{r}_{m-1}\}, \; \bm{r}_0 = \bm{r},\bm{r}_1=\bm{L}\bm{r}_0,\, \\ 
    \text{and} \, \bm{r}_l = \bm{L}\bm{r}_{l-1} 
    + \bm{B}\bm{r}_{l-2}\, \text{for}\, l \geq 2,
\end{multline}
where we define the coefficient matrices $\bm{L} = -\bm{A}(\bar{\omega})^{-1}\bm{A}'(\bar{\omega})$ and $\bm{B} = -\bm{A}(\bar{\omega})^{-1} i \bm{M}$. $\bm{r}_0,\bm{r}_1,\dots,\bm{r}_{m-1}$ is a second-order Krylov sequence and $m$ is the number of matched moments. For the derivative of $\bm{A}(\bar{\omega})$ with respect to the expansion frequency $\bar{\omega}$ we find $\bm{A}'(\bar{\omega}) = -2 \bar{\omega} \bm{M} + i \beta /c \bm{D}$. 
To build the actual projection matrix $\bm{V}$ to  subspace \eqref{eqn:soks}, we adopt the adaptive order rational Arnoldi-type method (AORA) \cite{LeeChuFe06} as proposed in \cite{Bodendiek2013} to the second order case analogous to the SOAR procedure. In this way we obtain an adaptive order second-order rational Arnoldi-type method. This ensures orthonormality in the columns of $\bm{V}$ while at the same time several expansion frequencies can be used. A central component of the method is a modified Gram-Schmidt orthogonalization. Repeating the orthogonalization once can improve the numerical stability of the method, as demonstrated in \cite{gramschmidt2002}.
With the projection matrix $\bm{V}$ from the second order rational Arnoldi method, we obtain the following for the reduced order system:

\begin{equation}
    (-k^2 \bm{M}_r - i k \beta  \bm{D}_r + \bm{S}_r) \bm{u}_r(\omega) = \bm{F}_r(\bar{\omega}),
\end{equation}
where $\bm{M}_r = \bm{V}^{*}\bm{M}\bm{V} \in \mathbb{C}^{r\times r}$ and similar expressions hold for $\bm{S}$ and $\bm{D}$. The error of this moment matching ROM method depends on the number of matched moments $m$ and the choice of the expansion frequencie(s) $\bar{\omega}$. The transfer function of the ROM matches the Taylor series coefficients up to order $m$ of its full order counterpart, expanded around $\bar{\omega}$. 


\section{Statistical Reduced Order Models}\label{sec4}
In this section we introduce a statistical model based on the reduced order framework of Section \ref{sec3}. We first show, how a ROM can be used to efficiently carry out uncertainty forward propagation, followed by a presentation of the statFEM and statROM procedure, respectively.

\subsection{Forward Problem}
\label{Sec:ForwardProblem}
The statFEM approach \cite{girolami2021statistical} is in its core based on GPs, which form a class of Bayesian non-parametric models that are well established in statistics and machine learning \cite{murphy2012}. In a first step, a GP model over the FE solution is formed, which is obtained by propagating uncertainty in material parameters, the geometry or excitations through the FE model. 
Here, to account for the uncertainty both in the material parameter and the right-hand-side, a ROM is used. Then, in a second step, data can be used for updating the state and for reducing the prediction uncertainty. In this way, statFEM makes use of GPs to fuse a solution from a FE solver with data. When modelling the state as a GP, normally distributed model and data errors give rise to a normally distributed posterior, hence Bayesian updating can be carried out in closed form. This in turn is convenient for a monitoring perspective, where fast sampling free updating procedures are desired.  

 %
%
%

In our model, the uncertainty is in the simplest case assumed to be present in the inhomogeneous Neumann boundary condition. More precisely, we assume $g$ to be a random field 
\begin{equation}
    g(\bm{x}) \sim \mathcal{GP}(\mu_{g}(\bm{x}), c_g(\bm{x},\bm{x}')),
\end{equation}
with the covariance defined via the Mat\'ern kernel
\begin{equation}
     c_g(\bm{x},\bm{x}') = c_{\text{Mat\'ern}}(\bm{x},\bm{x}') = \sigma^2 \frac{2^{1-\nu}}{\Gamma(\nu)}  \left( \frac{\sqrt{2\nu}\| \bm{x} - \bm{x}' \|}{l}  \right)^{\nu} K_{\nu}  \left(  \frac{\sqrt{2\nu}\| \bm{x} - \bm{x}' \|}{l}   \right),
\label{eqn:MaternKernel}
\end{equation}
with the scaling parameter $\sigma \in \mathbb{R}^+$, $\Gamma(\nu)$ the Gamma function and $K_{\nu}$ a modified Bessel function \cite[p.84 ff.]{abramowitz2013}, \cite{rasmussen2006}. Moreover, $l \in \mathbb{R}^+$ is the lengthscale parameter and $\nu\in \mathbb{R}^+$ controls the smoothness of the kernel.
For $\nu \to \infty$  (\ref{eqn:MaternKernel}) becomes the squared exponential kernel. The Mat\'ern kernel is used here because of its flexibility to represent variable smoothness. 

We employ the following discrete representations of the mean and covariance function of the corresponding forcing term $\bm{f}_g$, 
\begin{equation}
    (\boldsymbol{\mu}_{g})_i = \int_{\Gamma_{\mathrm{N},g}} \mu_g \phi_i^* \mathrm{d} x,
    \label{eqn:rhsDiscret}
\end{equation}
where $\phi_i(\bm{x})$ denotes the nodal FE basis function associated to node $\bm{x}_i$ of the mesh, as well as 
%
%
\begin{multline}
(\bm{C}_{g})_{i,j} =  \int_{\Gamma_{\mathrm{N},g}} \int_{\Gamma_{\mathrm{N},g}} \phi_i(\bm{x})  c_g(\bm{x},\bm{x}') \phi_j^*(\bm{x}')    \,\mathrm{d}x   \,\mathrm{d}x' \\
\approx \left( \int_{\Gamma_{\mathrm{N},g}} \phi_i    \,\mathrm{d}x \right) c_g(\bm{x}_i,\bm{x}_j) \left( \int_{\Gamma_{\mathrm{N},g}}    \phi_j^*    \,\mathrm{d}x \right).
\label{eqn:covMatDerivation}
\end{multline}
In \eqref{eqn:covMatDerivation}, we follow \cite{girolami2021statistical} and interpolate the covariance function with the FE basis functions. Mass lumping then leads to the approximation in \eqref{eqn:covMatDerivation} that still will be denoted as $(\bm{C}_{g})_{i,j}$ in the following. The associated approximation error is expected to be small, at the same time, the implementation largely simplifies. We refer to \cite{girolami2021statistical} and the references therein for a discussion. 
In view of the linear dependence of $\bm{f}_g$ on $g$, the source term can be represented by a multivariate Gaussian
\begin{equation}
    \bm{f}_g \sim \mathcal{N}(\boldsymbol{\mu}_{g}, \bm{C}_{g}).
    \label{eqn:forcingNormal}
\end{equation}

In addition to $\bm{f}_g$, we consider a volume source term on the right-hand-side, corresponding to $f$ in \eqref{eq:model_strong}. This term is model as a Gaussian random field
\begin{equation}
    f(\bm{x}) \sim \mathcal{GP}(\mu_f(\bm{x}), c_f(\bm{x},\bm{x}'))
\end{equation}
as well, with $c_f(\bm{x},\bm{x}')$ defined via the Matérn kernel. 
Applying the same discretization principle as before, cf. \eqref{eqn:rhsDiscret} - \eqref{eqn:forcingNormal}, we obtain
\begin{equation}
    \bm{f} \sim \mathcal{N}(\boldsymbol{\mu}_f, \bm{C}_{f}).
    \label{eqn:forcing}
\end{equation}
We further consider uncertainty in the material parameter, given as 
\begin{equation}
    \kappa(\bm{x})  \approx \exp( \mu_{\kappa}(\bm{x}) + \sum_{i=1}^{M} \sqrt{\lambda_i} \Psi_i(\bm{x}) \xi_i),
\end{equation}
obtained as the Karhunen-Lo\`eve expansion of the covariance $c_{\ln(\kappa)}$,
\noindent where $\xi_i \sim \mathcal{N}(0,1)$ are independent and $(\lambda_i,\bm{\Psi}_i)$ are the eigenpairs of the covariance operator induced by $c_{\ln(\kappa)}$. The expansion has been truncated after $k$ terms according to some truncation criterion, e.g., the fraction of explained variance.

\begin{remark}
    Complex-valued Gaussian random vectors can be characterized by the mean, covariance and pseudo-covariance. The pseudo-covariance can play an important role in GP regression, see for instance \cite{bect2023rational}. Here, we treat the real and imaginary parts independently with individual Gaussian distributions based on the same covariance, which corresponds to setting the pseudo-covariance to zero. This is referred to as circular case in the signal processing literature. 
\end{remark}

Now the system matrix depends on the Gaussian vector $\bm{\xi}$ and the stochastic parametric FOM reads 
\begin{equation}
    \label{eq:FON_with_split_RHS}
        \bm{A}(\omega,\bm{\xi}(\theta)) \bm{u}(\omega,\theta) = \bm{F}(\omega,\theta),
\end{equation}
where $\bm{F}(\omega,\theta) = \bm{f}(\omega,\theta) + \bm{f}_g(\omega,\theta)$
and $\bm{\xi},\bm{f}(\omega,\cdot),\bm{f}_g(\omega,\cdot)$ are independent random vectors on the same probability space, collected as $\bm{\eta} = (\bm{\xi}^\top,\bm{f}(\omega,\cdot)^\top,\bm{f}_g(\omega,\cdot)^\top)^\top \in \mathbb{R}^M \times \mathbb{C}^{2N}$.


In the original statFEM paper \cite{girolami2021statistical}, the authors employ a perturbation approach to propagate forward probabilistic material coefficients that appear in the FEM system matrix. We, instead, apply a quasi Monte Carlo (QMC) method \cite{MOROKOFF1995} to approximate the prior mean and covariance of $\bm{u}$. To this end,
let $\tilde{\bm{\eta}}^{(i)} \in [0,1]^{M+2N} \ i = 1,\ldots, Q$, denote a sequence based on digital nets with random shifting \cite{Kuo2016}. Since all elements of $\bm{\eta}$ are normally distributed, we use the transformation 
\begin{equation}
    \bm{\eta}^{(i)} = 
    \left (
        \begin{array}{c}
            \bm{\xi}^{(i)} \\ \bm{f}^{(i)}(\omega) \\\bm{f}_g^{(i)}(\omega)
        \end{array}
    \right)
    = 
    \left(
        \begin{array}{c}
            \bm{0} \\ \bm{\mu}_f(\omega) \\\bm{\mu}_g(\omega)
        \end{array}
    \right)
    + 
    \left(
        \begin{array}{ccc}
            \bm{I} & \bm{0} & \bm{0} \\
            \bm{0} & \bm{C}_f^{1/2} & \bm{0} \\
            \bm{0} & \bm{0} & \bm{C}_g^{1/2}
        \end{array}
    \right)
    \Phi^{-1}(\tilde{\bm{\eta}}^{(i)})
    \label{eq:quasiMonteCarlo}
\end{equation}
 where $\Phi$ denotes the CDF of a standard Gaussian random variable, and the inverse is applied component-wise in \eqref{eq:quasiMonteCarlo}. Note that, because of the linearity of $\bm{A}$, the randomness in the right-hand-side could be propagated analytically to the solution. However, we employ the QMC sampling approach in the same way for all uncertain inputs, for simplicity.  We construct for each sample $\bm{\xi}^{(i)}$ the ROM projection matrix $\bm{V}^{(i)}$ as defined in Section \ref{sec3}. 
Then, we can efficiently solve the system 
\begin{equation}
    \label{eq:QMC_full}
        (\bm{V}^{(i)})^* \bm{A}(\omega,\bm{\xi}^{(i)}) \bm{V}^{(i)} \bm{u}_r^{(i)}(\omega) = (\bm{V}^{(i)})^* \bm{F}^{(i)}(\omega),
\end{equation}
for a large number of frequency points $\omega$ and right-hand-sides and repeat the procedure for different $\bm{\xi}^{(i)}$. 


Compared to the QMC method, a perturbation approach or surrogate methods, such as generalized polynomial chaos (gPC) \cite{Xiu2002}, would be faster for a small numbers of parameters. The QMC approach is however well-suited for handling the moderately large amount of uncertain inputs, while yielding a higher order convergence compared to pure MC.  



Given the the ROM projection matrices $\bm{V}^{(i)}, i=1,\ldots,Q$, for every query point $\omega$, we compute a Gaussian approximation as 
\begin{equation}
    \bm{u}(\omega) \approx \bm{u}_{Q,r}(\omega) \sim \mathcal{N}(\bm{\mu}_{\bm{u};Q,r}(\omega), \bm{C}_{\bm{u};Q,r}(\omega)),
    \label{eq:priormeanrom}
\end{equation}
where we first solve for $\bm{u}_r^{(i)}(\omega), i=1,\ldots,Q$ using \eqref{eq:QMC_full} and then build the standard Monte Carlo mean and covariance approximation 
\begin{align}
    \bm{\mu}_{\bm{u}}(\omega) &\approx \bm{\mu}_{\bm{u};Q,r}(\omega) = \frac{1}{Q} \sum_{i=1}^Q \bm{V}^{(i)} \bm{u}^{(i)}(\omega) \\
    \bm{C}_{\bm{u}}(\omega) &\approx \bm{C}_{\bm{u};Q,r}(\omega)  \notag \\
    &=\frac{1}{Q-1} \sum_{i=1}^Q \left(\bm{V}^{(i)} \bm{u}^{(i)}(\omega) - \bm{\mu}_{\bm{u};Q,r}(\omega)\right)\left(\bm{V}^{(i)} \bm{u}^{(i)}(\omega) - \bm{\mu}_{\bm{u};Q,r}(\omega)\right)^*.
\end{align}
A sample for $\bm{u}_{Q,r}(\omega)$ is then generated in the standard way as $\bm{\mu}_{\bm{u};Q,r}(\omega) + \bm{C}_{\bm{u};Q,r}^{1/2}(\omega) \bm{\chi}$,  with $\bm{\chi} \sim \mathcal{N}(\bm{0},\bm{I})$ of dimension $N$. In the following, to simplify the notation, we will denote the QMC-ROM Gaussian approximation as $\tilde{\bm{u}} = \bm{u}_{Q,r}$ and drop the dependence on $\omega$ in the next subsection.
%

\subsection{statFEM}

The statFEM, introduced in \cite{girolami2021statistical}, is a method to assimilate data with a stochastic FE-based prior model. Contrary to traditional Bayesian parameter estimation methods, which infer an unknown model parameter, statFEM updates the \emph{state}, i.e., the FE-solution. The statFEM relies on GP models for the state, the data and the unknown model error, which are connected through the statistical generating model 
\begin{equation}
\bm{y} = \bm{z} + \bm{e} = \rho \bm{P} \bm{u} + \bm{d} + \bm{e}.
\label{eqn:statGen}
\end{equation}
In \eqref{eqn:statGen}, the data $\bm{y} \in \mathbb{C}^{n_y}$ are decomposed into the true but unknown states $\bm{z} \in \mathbb{C}^{n_y}$, at the sensor locations $\check{\bm{x}}_{i}, i=1,\ldots,n_y$ and the data error $\bm{e}\in \mathbb{C}^{n_y}$. The true state $\bm{z}$ can further be expressed as a combination of the FE prediction $\bm{u} \in \mathbb{C}^{N}$, scaled by an unknown factor $\rho$ and projected onto the sensor locations with the projection matrix $\bm{P} \in \mathbb{R}^{n_y \times N}$, and a statistical representation of the model error $\bm{d} \in \mathbb{C}^{n_y}$. According to \cite{girolami2021statistical}, the projection matrix $\bm{P}$ is computed by evaluating the FE basis functions at the sensor locations. Hence, to use statFEM with commercial FE codes, these at least need to provide information about the used basis functions or provide them explicitly. 

For the time being, we assume that the prior $p(\bm{u}) \sim \mathcal{N}(\bm{\mu}_{\bm{u}},\bm{C}_{\bm{u}})$ is fully available and postpone the discussion about numerical approximation errors. Then, the model discrepancy is given by the difference between the exact model response and the true state at the sensor locations. This is essentially a deterministic quantity, that can be reduced by improving the model. The choice of modelling $\bm{d}$ in a statistical way is made out of convenience, but also to improve the data assimilation accuracy, see \cite{calvetti2018iterative} for example. More precisely, following \cite{girolami2021statistical} we set
\begin{equation}
\bm{d} \sim p(\bm{d} | \bm{\alpha}) = \mathcal{N}(\bm{0},\bm{C_{\bm{d}}}(\bm{\alpha}))
\end{equation}
with $\bm{\alpha}$ introduced as the vector of hyperparameters of the chosen kernel function. Here, the Mat\'ern kernel \eqref{eqn:MaternKernel} is also chosen to represent the covariance for the model error, hence, $c_d(\bm{x},\bm{x}';\bm{\alpha}) = c_{\text{Mat\'ern}}(\bm{x},\bm{x}';\bm{\alpha})$ where we now explicitly account for the hyperparameters $\bm{\alpha} = (\sigma,l)$. The covariance $\bm{C_{\bm{d}}}(\bm{\alpha})$ is then obtained by interpolating this kernel at the sensor locations as
\begin{equation}
    (\bm{C_{\bm{d}}}(\bm{\alpha}))_{ij} = c_d(\check{\bm{x}}_i,\check{\bm{x}}_j;\bm{\alpha}).
\end{equation}

The hyperparameters are estimated by maximising the marginal likelihood, i.e. the likelihood of the data given the hyperparameters $p(\bm{\bm{y}|\bm{\alpha}})$. This is detailed in Section~\ref{sec:statROM}. Please note, that the scale of the model error will be reflected in the parameter $\sigma$ of \eqref{eqn:MaternKernel}. Hence, an improvement of the model will result in a smaller estimated value of $\sigma$ and hence a smaller uncertainty band for the model error, see also \cite{narouie2022}. Evaluating the kernel function at the measurement positions, the GP is represented by a multivariate Gaussian with a covariance matrix $\bm{C_d} \in \mathbb{R}^{n_y \times n_y}$.

The assumption of normally distributed data is ubiquitous and often justifiable for specific sensors and measurement equipment. Hence, $\bm{e}$ is modelled as a multivariate Gaussian 
$\bm{e} \sim p(\bm{e}) = \mathcal{N}(\bm{0}, \bm{C_e})$
with zero mean and the covariance matrix $\bm{C_e} = \sigma_e^2 \bm{I}$.

After the hyperparameters $(\bm{\alpha},\rho)$ are learned based on the given data $\bm{y}$, the prior state is conditioned on the same observation data, following the empirical Bayes procedure \cite{Casella1985}. This leads to an optimal set of hyperparameters and a posterior state which accounts for both FE solution, data and model uncertainty.

Despite being very recent, the method has already been adopted in more practical scenarios and has also been studied and developed further to account for different use cases. First elements of error analysis are provided in \cite{Papandreou2021,KARVONEN2022}. In \cite{DUFFIN2022lowrank}, numerical difficulties which arise when scaling the method from academic examples to bigger systems are analysed and overcome by introducing a low-rank approach to keep memory usage small and the linear algebra tractable. Relying on Langevin dynamics, an efficient way to solve the statFEM forward problem, without using full PDE solves, is introduced in \cite{akyildiz2021}. An adaptation of statFEM for non-linear and time dependent problems is introduced in \cite{Duffin2020}. Furthermore, an application in civil engineering is demonstrated in \cite{febrianto_butler_girolami_cirak_2022}, where statFEM is used to enable digital-twinning of a structure. An application in solid mechanics has been proposed in \cite{narouie2022}, where a prior derived from a simple material model is conditioned on data which stem from a more complex material model. Therein, the forward problem is solved using a polynomial chaos approach. 

Before we adapt the statFEM data generating model to reflect the usage of approximate ROMs, we recall the conditioning procedure formulated in \cite{girolami2021statistical}.
Let $\bm{Y}$ denote the matrix containing $n_o$ readings $\bm{y}_i, i=1\ldots,n_o$ of the sensor vector stacked column-wise. With \eqref{eqn:statGen}, the posterior moments read
\begin{equation}
\bm{\mu}_{\bm{u}|\bm{Y}} = \bm{C}_{\bm{u}|\bm{Y}} \left(   \rho \bm{P}^T  (\bm{C}_{\bm{d}} + \bm{C}_{\bm{e}})^{-1}  \sum_{i=1}^{n_o}\bm{y}_i  +  \bm{C}_{\bm{u}}^{-1}  \bm{\mu}_{\bm{u}}   \right)
\end{equation}
and
\begin{equation}
\bm{C}_{\bm{u}|\bm{Y}} = \left( \rho^2 n_o \bm{P}^T   (\bm{C}_{\bm{d}} + \bm{C}_{\bm{e}})^{-1}  \bm{P}  +  \bm{C}_{\bm{u}}^{-1}    \right)^{-1} 
\end{equation}
with $n_o \in \mathbb{N}^{+}$ the number of observations per sensor and
\begin{equation}
\label{eqn:MultObs}
p(\bm{Y}|\bm{u}) = p(\bm{y}_1|\bm{u}) p(\bm{y}_2|\bm{u})\ldots p(\bm{y}_{n_o}|\bm{u}) = \prod_{i=1}^{n_o}p(\bm{y}_i|\bm{u}),
\end{equation}
where we assumed independence of the different readings. From the posterior and the inferred model error, the inferred true response can be found as
\begin{equation}
p(\bm{z}|\bm{Y}) = \mathcal{N}(\rho\bm{P}\bm{\mu}_{\bm{u}|\bm{Y}}, \rho^2 \bm{P}\bm{C}_{\bm{u}|\bm{Y}}\bm{P}^T + \bm{C}_{\bm{d}}).
\label{eqn:trueProcEstEq}
\end{equation}
Finally, the predicted data distribution at new sensor locations, with $\hat{\bm{P}} \in \mathbb{R}^{n_{\hat{y}} \times N}$, reads
\begin{equation}
p(\hat{\bm{y}}|\bm{Y}) = \mathcal{N}(\rho\hat{\bm{P}}\bm{\mu}_{\bm{u}|\bm{Y}}, \rho^2 \hat{\bm{P}}\bm{C}_{\bm{u}|\bm{Y}}\hat{\bm{P}}^T + \hat{\bm{C}}_d + \hat{\bm{C}}_e) 
\end{equation}
for the $n_{\hat{y}}$ DOFs where no observations are made. These findings shall now be adapted to the case of a ROM prior with a controllable approximation error.

\subsection{ROM error estimation}
\label{sec:RomErrorAdj}
Contrary to the FE-to-reality error, the error introduced by the ROM is easier to assess and it is desirable to include an estimate of it into the data generating model. The ROM approximation error can be estimated in an expensive but accurate way by sampling and comparing the FE solution and ROM prediction at each sample point. 
The so-obtained ROM error samples are $\bm{d}_{r}^{(i)}(\omega) = \bm{u}^{(i)}(\omega) - \bm{V}^{(i)} \bm{u}_r^{(i)}(\omega)$, from which the mean and variance can be estimated. This procedure is of limited use here, however, since it would be better to use $\bm{u}^{(i)}$ in the data assimilation step right-away. Hence, we are interested in finding a different error indicator, which is accurate enough to improve the state estimation and at the same time the additional cost of computing the error indicator must be moderate. 

We assume that the error is smooth enough over the computational domain so that again a GP model can be used to represent it. In particular, we estimate the ROM error at a few randomly chosen points and condition a GP to obtain an approximation over the entire domain. It is important to quantify the ROM error over the entire domain, not only at the sensor locations, to improve also the prediction of the true response $\bm{z}|\bm{Y}$ and $\hat{\bm{y}}|\bm{Y}$. Since, in this approach, the error is only needed at a few ($n_d$) points, we can employ adjoint methods. We follow an approach similar to the one presented in \cite{Rouse2022}, where the authors estimated the full FEM error field, contrary to the full ROM error field as desired here. Consider the solution $\bm{u}^{(i)}$ and the adjoint equation
\begin{equation}
    (\bm{A}(\omega,\bm{\xi}^{(i)}))^* \bm{q}_\ell^{(i)}(\omega) = \bm{p}_\ell,
\label{eq:adjoint}
\end{equation}
where $l = 1,\ldots, n_d$ and $\bm{p}_\ell^* \bm{u}(\omega) \approx u(\omega,\bm{x}_\ell)$. Then, the error related to $\bm{x}_\ell$ can be rewritten as 
\begin{align}
d_{r,\ell}^{(i)}(\omega) &= \bm{p}_\ell^* \bm{u}^{(i)}(\omega) - \bm{p}_\ell^* \bm{V}^{(i)} \bm{u}_r^{(i)}(\omega) \\
& =\left((\bm{A}(\omega,\bm{\xi}^{(i)}))^* \bm{q}_\ell^{(i)}(\omega)\right)^* \left( \bm{u}^{(i)}(\omega) - \bm{V}^{(i)}\bm{u}_r^{(i)}(\omega) \right) \\
& = (\bm{q}_\ell^{(i)}(\omega))^*  \bm{A}(\omega,\bm{\xi}^{(i)}) \left( \bm{u}^{(i)}(\omega) - \bm{V}^{(i)}\bm{u}_r^{(i)}(\omega) \right) \\
& = (\bm{q}_\ell^{(i)}(\omega))^* \left( \bm{F}^{(i)}(\omega) - \bm{A}(\omega,\bm{\xi}^{(i)})\bm{V}^{(i)} \bm{u}_r^{(i)}(\omega) \right).
\end{align}
This, however, still involves solving \eqref{eq:adjoint}, i.e., a full order system of equations. To reduce the computational effort, we again construct a ROM as
\begin{equation}
    (\bm{A}_r(\omega,\bm{\xi}^{(i)}))^* \bm{q}_{r,\ell}^{(i)}(\omega) = \bm{p}_{r,\ell}^{(i)},
\label{eq:adjoint_rom}
\end{equation}
where $(\bm{A}_r(\omega,\bm{\xi}^{(i)}))^* = (\bm{V}_{\mathrm{adj},\ell}^{(i)})^{*} (\bm{A}(\omega,\bm{\xi}^{(i)}))^* \bm{V}_{\mathrm{adj},\ell}^{(i)}$,  $\bm{p}_{r,\ell}^{(i)} = (\bm{V}_{\mathrm{adj},\ell}^{(i)})^*\bm{p}_\ell$ and $\bm{q}_{\ell}^{(i)}(\omega) = \bm{V}_{\mathrm{adj},\ell}^{(i)} \bm{q}_{r,\ell}^{(i)}(\omega)$. Please note that the construction of $\bm{V}_{\mathrm{adj},\ell}^{(i)} \neq \bm{V}^{(i)}$ has to be carried out additionally in the offline phase. However, running the moment matching algorithm again for $i = 1,\ldots, I$ is very cheap as we can reuse the LU decomposition of $\bm{A}(\omega,\bm{\xi}^{(i)})$ which had to be computed already to find the ROM prior. 

\begin{remark}
\label{rmk:AORGA}
Note that for multiple right-hand-sides, we can use AORGA (Adaptive order rational global Arnoldi) \cite{bodendiek2013Diss,JBILOU199949} instead of AORA, which would decrease the number of needed projection matrices. Instead of a right hand side vector, AORGA expects a right hand side matrix to find a projection basis.
\end{remark}

We pick a problem-dependent (small) number of points $\bm{X}= [\bm{x}_1,\ldots,\bm{x}_{n_\mathrm{adj}}]$ throughout the domain and compute the adjoint error estimate $\bm{d}_{r,\mathrm{adj}}$, i.e., the mean over the QMC sample.
Then, a zero mean GP prior with Matérn kernel $c_{d_r}(\bm{x},\bm{x}')$
\begin{equation}
    d_r(\bm{x}) \sim \mathcal{GP}(0,c_{d_r}(\bm{x},\bm{x}'))
\end{equation}
is conditioned on the vector of pointwise estimated errors $\bm{d}_{r,\mathrm{adj}}$ to find the posterior at the points set $\hat{\bm{X}}$
\begin{equation}
    \hat{\bm{d}}_{r} | \bm{d}_{r,\mathrm{adj}} \sim \mathcal{N}(\bm{\mu}_{\hat{\bm{d}}_r},\bm{C}_{\hat{\bm{d}}_r}),
    \label{eq:rom_error_gp}
\end{equation}
where
\begin{align}
    \bm{\mu}_{\hat{\bm{d}}_r} &= \bm{C}_{\bm{d}_{r};\hat{\bm{X}},\bm{X}} \bm{C}_{\bm{d}_{r};\bm{X},\bm{X}}^{-1} \bm{d}_{r,\mathrm{adj}} \\
    \bm{C}_{\hat{\bm{d}}_r} &= \bm{C}_{\bm{d}_{r};\hat{\bm{X}},\hat{\bm{X}}} - \bm{C}_{\bm{d}_{r};\hat{\bm{X}},\bm{X}} (\bm{C}_{\bm{d}_{r};\bm{X},\bm{X}}^{-1} +\bm{I} \bm{\sigma}_\mathrm{adj}^2)\bm{C}_{\bm{d}_{r};\bm{X},\hat{\bm{X}}}
\end{align}
and $\bm{C}_{\bm{d}_{r};\hat{\bm{X}},\bm{X}}$ denotes the prior covariance evaluated at the point sets $\hat{\bm{X}}$ and $\bm{X}$, respectively and a similar convention applies for the other matrices. Also, $\bm{\sigma}_\mathrm{adj}^2$ is the vector of variances obtained while sampling $\bm{d}_{r,\mathrm{adj}}$.

In the next sections, this ROM approximation error indicator will be included into the statFEM posterior update and it will be shown that this can help improve the posterior mean estimate.

\begin{remark}
If the problem is subject only to a stochastic right hand side $\bm{F}$ it is possible to compute a prior for $\bm{d}_r$ in closed form.
In turn, the number of required projection matrices can be reduced. Let $\bm{F}_r \sim \mathcal{N}(\bm{\mu}_{\bm{F}_r},\bm{C}_{{\bm{F}}_r})$ be the reduced stochastic right-hand-side. We propagate its distribution to $\bm{u}_r$ via 
\begin{equation}
\bm{A}_r(\omega) \bm{u}_r(\omega) = \bm{F}_r.
\end{equation}
First, the mean can be computed in closed form as $\bm{\mu}_{\bm{u}_r}(\omega) = \bm{A}_r(\omega)^{-1} \bm{\mu}_{{\bm{F}}_r}$. Moreover, 
\begin{equation}
\bm{q}_\ell^*  \bm{F}  \sim \mathcal{N}(\bm{q}_\ell^*\bm{\mu}_{\bm{F}}, \bm{q}_\ell^* \bm{C}_{\bm{F}} \bm{q}_\ell)
\end{equation}
and 
\begin{equation}
\bm{q}_\ell^* \bm{A}(\omega)\bm{V} \bm{u}_r(\omega) \sim \mathcal{N}(\bm{q}_\ell^*\bm{B}(\omega)\bm{\mu}_{{\bm{F}}_r}, (\bm{q}_\ell^* \bm{B}(\omega))^* \bm{V}^*\bm{C}_{\bm{F}}\bm{V} (\bm{q}_\ell^* \bm{B}(\omega)))
\end{equation}
with $\bm{B}=  \bm{A}\bm{V} \bm{A}_r^{-1}$. Please note that $\bm{V}$ is constructed with a deterministic $\bm{A}$ and a specific realization of $\bm{F}$ - as before - and hence, is deterministic in this case. Therefore, we obtain the following prior for the ROM-error 
\begin{equation}
d_{\ell}(\omega) \sim \mathcal{N}(\bm{q}_\ell^*(\bm{V} - \bm{B}(\omega))\bm{\mu}_{\bm{f}_r},\bm{q}_\ell^* \bm{C}_{\bm{f}} \bm{q}_\ell + (\bm{q}_\ell^* \bm{B}(\omega))^* \bm{V}^*\bm{C}_f\bm{V} (\bm{q}_\ell^* \bm{B}(\omega))).
\label{eq:adj_closedform}
\end{equation}
\end{remark}


\subsection{statROM}
\label{sec:statROM}
%
%
A standard statFEM posterior update can of course be conducted, with the prior FE solution replaced with a ROM. However, the error model employed in \eqref{eqn:statGen} is rather crude and the additional ROM approximation error would simply inflate $\bm{C}_{\bm{d}}$. In the last section, we have formulated a way to approximate the ROM error for every point in the domain. Knowing more about the numerical error should, intuitively, lead to a better solution, i.e., a sharper estimated model error and, for the predictive observation density, a posterior mean closer to the observed data. Therefore, in this section, the statistical generating model and the posterior update equations are modified to account for the ROM approximation error. We label the resulting updating procedure as statistical reduced order modelling (statROM).

In analogy to the statFEM case, beginning with the statFEM statistical generating model \eqref{eqn:statGen}, we expand the FE solution term as
\begin{equation}
\bm{y} = \rho \bm{P}( \bm{u} -\bm{V} \bm{u}_r + \bm{V} \bm{u}_r)+ \bm{d} + \bm{e}, 
\end{equation}
which yields
\begin{align}
\label{eqn:datamodel}
\bm{y} &= \rho \bm{P} \bm{V} \bm{u}_r+ \rho \bm{P}(\bm{u}-\bm{V} \bm{u}_r) + \bm{d} + \bm{e} \\
\bm{y} &= \rho \bm{P} \bm{V} \bm{u}_r+ \rho \bm{P}\bm{d}_r + \bm{d} + \bm{e}
\end{align}
with $\bm{d}_r \in \mathbb{C}^{N}$ representing the ROM approximation error, which is obtained by interpolating the ROM error GP at the FE mesh nodes. 
Unlike the model error $\bm{d}$, the ROM error indicator involves a non-zero mean. Hence, a bias induced by the ROM can directly be corrected and the ROM approximation error does not only inflate the variance. The statistical generating model is now used to derive the posterior update equations.

To condition the ROM prior on the data, at first, the joint distribution is constructed as
\[ 
p(\tilde{\bm{u}}, \bm{y}) = \mathcal{N} \left( \begin{pmatrix}
                                          \bm{\mu}_{\tilde{\bm{u}}} \\
                                           \bm{\mu}_{\bm{y}}
                                         \end{pmatrix} , \; \begin{pmatrix}
                                                                  \mathbb{E}\left( ( \tilde{\bm{u}} - \bm{\mu}_{\tilde{\bm{u}}})  ( \tilde{\bm{u}} - \bm{\mu}_{\tilde{\bm{u}}})^*  \right)          & \mathbb{E}\left(  ( \tilde{\bm{u}} - \bm{\mu}_{\tilde{\bm{u}}})  ( \bm{y} - \bm{\mu}_{\bm{y}})^*  \right)   \\
                                                                  \mathbb{E}\left( ( \bm{y} - \bm{\mu}_{\bm{y}})  ( \tilde{\bm{u}} - \bm{\mu}_{\tilde{\bm{u}}}) ^*  \right)             &\mathbb{E}\left( ( \bm{y} - \bm{\mu}_{\bm{y}})  ( \bm{y} - \bm{\mu}_{\bm{y}})^*  \right) 
                                                                 \end{pmatrix} \right) \;.
\]
%
Conditioning $\tilde{\bm{u}}$ on $\bm{y}$ now yields, using the Sherman-Morris-Woodbury identity, for the posterior mean
\begin{equation}
\begin{split}
\bm{\mu}_{\tilde{\bm{u}} | \bm{y}} = \bm{\mu}_{\tilde{\bm{u}}} + \rho \bm{C}_{\tilde{u}} \bm{P}^\top \left[ \rho^2 \bm{P} \bm{C}_{\tilde{\bm{u}}} \bm{P}^\top  + \bm{K}_{\bm{y}_r}  \right]^{-1} \left(  \bm{y} - \left( \rho \bm{P} \bm{\mu}_{\tilde{\bm{u}}} + \rho \bm{P}\bm{\mu}_{\bm{d}_r}  \right) \right)
\end{split}
\end{equation}
with $\bm{K}_{\bm{y}_r} = \rho^2 \bm{P}\bm{C}_{\bm{d}_r}\bm{P}^{\top}  + \bm{C}_{\bm{d}} + \bm{C}_{\bm{e}}$ and for the covariance
\begin{equation}
\bm{C}_{\tilde{\bm{u}}|\bm{y}} = \bm{C}_{\tilde{\bm{u}}} - \rho \bm{C}_{\tilde{\bm{u}}} \bm{P}^\top \left[ \rho^2 \bm{P} \bm{C}_{\tilde{\bm{u}}} \bm{P}^\top  + \bm{K}_{\bm{y}_r}  \right]^{-1} \rho \bm{P} \bm{C}_{\tilde{\bm{u}}}
\end{equation}
where we recall that $\tilde{\bm{u}} \sim \mathcal{N}(\bm{\mu}_{\tilde{\bm{u}}},\bm{C}_{\tilde{\bm{u}}})$ is the ROM-QMC prior solution. For multiple readings, i.e. $n_o > 1$, we obtain
\begin{equation}
\bm{\mu}_{\tilde{\bm{u}}|\bm{Y}} = \bm{\mu}_{\tilde{\bm{u}}} + \rho \bm{C}_{\tilde{\bm{u}}} \bm{P}^\top \left[ \rho^2 n_o \bm{P} \bm{C}_{\tilde{\bm{u}}} \bm{P}^\top  + \bm{K}_{\bm{y},r}  \right]^{-1} \left(  \textstyle\sum_{i=1}^{n_o}\bm{y}_i - \rho n_o\left(  \bm{P} \bm{\mu}_{\tilde{\bm{u}}} + \bm{P}\bm{\mu}_{\bm{d}_r}  \right) \right)
\label{eqn:statROMpostMean}
\end{equation}
and for the covariance
\begin{equation}
\bm{C}_{\tilde{\bm{u}}|\bm{Y}} = \bm{C}_{\tilde{\bm{u}}} - \rho n_o \bm{C}_{\tilde{\bm{u}}} \bm{P}^\top \left[ \rho^2 n_o \bm{P} \bm{C}_{\tilde{\bm{u}}} \bm{P}^\top  + \bm{K}_{\bm{y}_r}  \right]^{-1} \rho \bm{P} \bm{C}_{\tilde{\bm{u}}}.
\label{eqn:statROMpostCov}
\end{equation}
Hence, for the statROM posterior we have
\begin{equation}
    \tilde{\bm{u}} | \bm{Y} \sim \mathcal{N}(\bm{\mu}_{\tilde{\bm{u}} |\bm{Y}},\bm{C}_{\tilde{\bm{u}} | \bm{Y}}).
    \label{eqn:statROMpost}
\end{equation}
The equations can also be derived by directly using Bayes' law as shown in \cite{girolami2021statistical}.
Adapting the argumentation of \cite{girolami2021statistical} to the present setting, we can deduce that for a covariance term $\bm{K}_{\bm{y}_r}$ which is small in comparison to $\bm{C}_{\tilde{\bm{u}}}$, the posterior mean tends to $(\sum_{i=1}^{n_o}\bm{y}_i - n_o\bm{\mu}_{\bm{d}_r})/(\rho n_o)$ and the posterior covariance tends to $\bm{K}_{\bm{y}_r}/\rho^2$. For a large covariance term, the posterior is close to the prior.
The higher the number of observations $n_o$, the more the posterior covariance tends to zero and the mean to the empirical mean of the ROM-error corrected observations $(\sum_{i=1}^{n_o}\bm{y}_i - n_o\bm{\mu}_{\bm{d}_r})/n_o$.
\subsubsection{Hyperparameter Learning}
To learn the posterior hyperparameters from data, the marginal likelihood has to be derived for the statROM case, as well. For a single observation the following holds for the statROM data model:
\begin{multline}
\label{eqn:margLik}
p(\bm{y}_i \vert \bm{\alpha},\rho) = \frac{1}{(2\pi)^{n_y/2}\lvert \bm{K} \rvert }  \cdot \\
\cdot \exp \left(-\dfrac{1}{2} (\bm{y}_i - \rho \bm{P} \bm{\mu}_{\tilde{\bm{u}}} - \rho\bm{P}\bm{\mu}_{\bm{d}_r})^\top \bm{K}^{-1} (\bm{y}_i - \rho \bm{P}\bm{\mu}_{\tilde{\bm{u}}} - \rho \bm{P}\bm{\mu}_{\bm{d}_r})\right)
\end{multline}
with $n_y$ the number of sensors and
\begin{equation}
 \bm{K} = \rho^2 \bm{P} \bm{C}_{\tilde{\bm{u}}}\bm{P}^\top + \rho^2 \bm{P}\bm{C}_{\bm{d}_r}\bm{P}^{\top} + \bm{C}_{\bm{d}} + \bm{C}_{\bm{e}}
\end{equation}
the covariance of \eqref{eqn:datamodel}. Taking the natural logarithm yields
\begin{multline}
    \log p(\bm{y}_i\vert \bm{\alpha},\rho) = -\dfrac{n_y}{2} \log (2 \pi) - \dfrac{1}{2} \log \lvert \bm{K} \rvert \\
    - \dfrac{1}{2} (\bm{y}_i - \rho \bm{P}\bm{\mu}_{\tilde{\bm{u}}} - \rho\bm{P}\bm{\mu}_{\bm{d}_r})^\top \bm{K}^{-1} (\bm{y}_i - \rho \bm{\mu}_{\tilde{\bm{u}}} - \rho \bm{P}\bm{\mu}_{\bm{d}_r}).
\label{eqn:logLikROM}
\end{multline}
As \eqref{eqn:MultObs} holds, we obtain
\begin{equation}
\log p(\bm{Y}\vert \bm{\alpha},\rho) = \sum_{i=1}^{n_o} \log p(\bm{y}_i\vert \bm{\alpha},\rho).
\end{equation}
The marginal likelihood is minimized using a gradient based optimizer. In this case, L-BFGS-B \cite{Liu1989} is used. Notice that in \eqref{eqn:logLikROM} the misspecified ROM prior mean at the sensor locations $\rho \bm{P}\bm{\mu}_{\tilde{\bm{u}}}$ is corrected by the mean error estimate $\rho \bm{P}\bm{\mu}_{\bm{d}_r}$.

\paragraph{Predictive Observation Density}
As derived in \cite{girolami2021statistical} for the statFEM case, we find in analogy the statROM predictive observation density, i.e., an estimate for the real underlying process at sensor locations as
\begin{equation}
p(\bm{z}|\bm{Y}) = \mathcal{N}\left(\rho\bm{P}\bm{\mu}_{\tilde{\bm{u}}|\bm{Y}}  + \rho \bm{\mu}_{\bm{d}_r}, \rho^2\bm{C}_{\tilde{\bm{u}}|\bm{Y}}\bm{P}^T +\bm{C}_{\bm{d}_r} + \bm{C}_{\bm{d}} \right) 
\label{eqn:trueProcEstEqRom}
\end{equation}
and for points where no data has been observed $\hat{\bm{y}}$, as
\begin{equation}
    p(\hat{\bm{y}}\vert \bm{Y}) = \mathcal{N} \left( \rho \hat{\bm{P}} \bm{\mu}_{\tilde{\bm{u}}|\bm{Y}} +  \rho \bm{\mu}_{\bm{d}_r},  \hat{\bm{C}}_{\bm{d}_r}  + \hat{\bm{C}}_{\bm{d}} + \hat{\bm{C}}_{\bm{e}} + \rho^2 \hat{\bm{P}} \bm{C}_{\tilde{\bm{u}}|\bm{Y}} \hat{\bm{P}}^{\top} \right).
    \label{eqn:predObRom}
\end{equation}
By a slight abuse of notation, we denote with $\tilde{\bm{u}}_{Y;{d_r}}$ the entire predictive field. One may observe that the posterior mean and covariance are corrected by the model error and ROM error terms. Following the reasoning in \cite{girolami2021statistical}, the predictive observation density now tends, in contrast to the plain posterior, to $\bm{y}$ for the mean and to $2\bm{C}_d + \bm{C}_{d_r} + \bm{C}_e$ for the covariance, hence to the ground truth.

Algorithms \ref{algo1} and \ref{algo2} capture the implementation of the statROM workflow beginning from a full order reference model and ending at a numerically inexpensive predictive observation density. The process is divided into an offline and an online phase.


\begin{algorithm}[t]
\caption{Offline phase}\label{algo1}
\begin{algorithmic}[1]
\Require frequency expansion point(s) $\bar{\omega}$, ROM order $m$ 
\State generate $I$ QMC samples $\bm{\eta}^{(i)}$
\State define $n_\mathrm{adj}$ adjoint estimator training points $\bm{X}$
\ForAll{$\bm{\xi}^{(i)}, i=1,\ldots,I$}
    \State $\bm{M}^{(i)},\,\bm{S}^{(i)},\bm{D}^{(i)} = \mathrm{assembleSystem}(\bm{\xi}^{(i)})$ \Comment{FEM system assembly}
    \State $\bm{V}^{(i)}, \bm{L}^{(i)}, \bm{U}^{(i)} = \mathrm{AORA}(\bm{M}^{(i)},\bm{S}^{(i)},\bm{D}^{(i)},\bm{F}^{(i)},\bar{\omega}, m)$ \Comment{Primal ROM basis}
    \ForAll{$\bm{x} \in \bm{X}$}
        \State $\bm{V}_{\mathrm{adj},\ell}^{(i)} = \mathrm{AORA}((\bm{M}^{(i)})^*,(\bm{S}^{(i)})^*,(\bm{D}^{(i)})^*,\bm{p}_\ell,\bar{\omega}, m,\bm{L}^{(i)}, \bm{U}^{(i)})$ \Comment{Adjoint ROM basis}
    \EndFor
\EndFor
\end{algorithmic}
\end{algorithm}

\begin{algorithm}[t]
\caption{Online phase}\label{algo2}
\begin{algorithmic}[1]
\Require $\omega \in [\omega_{\text{min}},\omega_{\text{max}}]$, $\bar{\omega}$, $m$, $c$, $\bm{V}^{(i)}$, $\bm{V}_{\mathrm{adj},\ell}^{(i)}$
\Ensure $\omega$ non-resonant \Comment{(For an undamped} system)
\State $k= \omega/c$
\ForAll{$\bm{\xi}^{(i)}, i=1,\ldots,I$}
    \State $\bm{M}^{(i)},\,\bm{S}^{(i)},\bm{D}^{(i)} = \mathrm{assembleSystem}(\bm{\xi}^{(i)})$ \Comment{FEM system assembly}
    \State $\bm{A}^{(i)}(\omega) = (-k^2 \bm{M}^{(i)} - ik\beta\bm{D}^{(i)} + \bm{S}^{(i)})$
    \State $\bm{A}_{r}^{(i)} = (\bm{V}^{(i)})^*\bm{A}^{(i)}\bm{V}^{(i)}$
  \State $\bm{u}_{r}^{(i)} = (\bm{A}_{r}^{(i)})^{-1} \bm{F}_r^{(i)} $
    \State $\bm{d}_{r}^{(i)} = \mathrm{adjEst}(\bm{V}_{\mathrm{adj}}^{(i)},\bm{A}^{(i)},\bm{F}_r^{(i)} , \bm{u}_{r}^{(i)})$ \Comment{Adjoint error estimation \eqref{eq:rom_error_gp}}
    \EndFor
\State $\bm{U}_r = [\bm{u}_r^{(1)},\dots,\bm{u}_r^{(I)}]$
\State $\bm{D}_r = [\bm{d}_r^{(1)},\dots,\bm{d}_r^{(I)}]$
\State $\tilde{\bm{u}} \sim \mathcal{N}(\mathrm{mean}(\bm{V}\bm{U}_r),\mathrm{cov}(\bm{V}\bm{U}_r))$ \Comment{ROM prior \eqref{eq:priormeanrom}}
\State $\bm{d}_r \sim \mathcal{N}(\mathrm{mean}(\bm{D}_r ),\mathrm{cov}(\bm{D}_r))$ \Comment{define the ROM error, GP training}
\Require $\tilde{\bm{u}}, \bm{d}_r, \bm{Y}$, $\bm{P}$ \Comment{Measurement data and sensor locations}
\State $\tilde{\bm{u}}_{Y;{d_r}} = \mathrm{statROM}(\tilde{\bm{u}},\bm{d}_r,\bm{Y},\bm{P})$ \Comment{Compute posterior state \eqref{eqn:statROMpost}}

\end{algorithmic}
\end{algorithm}

\section{Numerical Examples}\label{sec5}
We consider two numerical examples to illustrate the new methods. The first one is a 1D example of the Helmholtz equation with random forcing and a random heterogeneous material coefficient. It is used to show the convergence properties of the proposed statROM approach compared to the standard approach without a ROM error estimator. The second example is more involved and deals with solving the Helmholtz equation in a 2D scattering scenario with a misspecified random forcing. For both settings, we compare the predictive observation density derived from the standard data model \eqref{eqn:trueProcEstEq} and from the proposed statROM data model \eqref{eqn:predObRom}. In both cases, the prior state is computed using the ROM. We denote the predictive state for the standard approach with $\tilde{\bm{u}}_{Y}$ and for the proposed statROM approach with $\tilde{\bm{u}}_{Y;{d_r}}$.

\subsection{One-dimensional problem}
    To illustrate the method and to show the convergence properties, we begin by investigating the Helmholtz equation in 1D. We consider an inhomogeneous Neumann boundary condition at $x_n = 0$ and no impedance boundary condition, hence, the problem is real-valued. We choose $f_g \sim \mathcal{N}(\frac{\pi^2}{50},0.02^2)$. At $x_n = 1$, a homogeneous Neumann boundary condition, i.e. a sound hard reflective boundary, is applied. The material parameter random field is parameterized with $\sigma_\kappa = \sqrt{5e-2},\;l_\kappa = 0.3$ for an exponential kernel function. The FE mesh consists of $100$ elements and linear Ansatz functions are employed. We first investigate the convergence of the ROM prior itself. Then we condition with the standard statFEM error model to obtain a reference, followed by a statROM update. Finally, we investigate the convergence of the predictive posterior solution towards the ground truth. 

\begin{figure}[ht]
\centering
\includegraphics[]{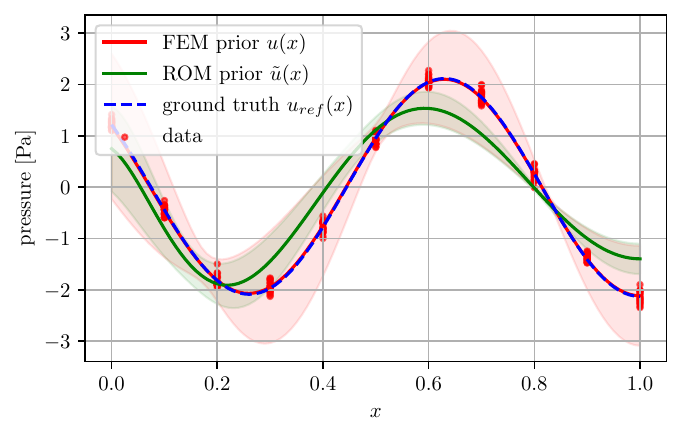}
\caption{Full order and ROM priors for the 1D problem. As the ROM is constructed with only $m=5$ moments matched around the expansion frequency, the approximation error is large. Observation data are collected on a finely discretized reference FE solution on $11$ equally spaced sensor locations. At each sensor, multiple noisy measurements (red dots) are recorded.} The shaded areas are the $2\sigma$ bands for the respective prior. The FEM prior mean and the ground truth align because no artificial model error is introduced. The only difference between both solutions is the finer discretisation of the ground truth.\label{fig:fig2}
\end{figure}

\begin{figure}[ht]
\centering
\includegraphics[]{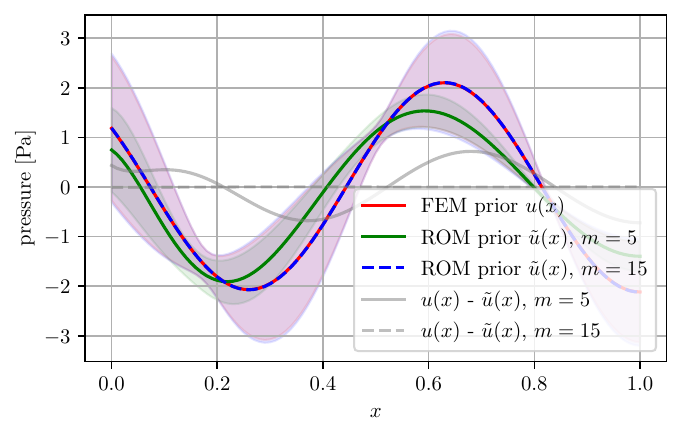}
\caption{Comparison of the quality of different ROM priors for the 1D problem. With $m=15$ moments matched, the approximation error is negligible compared to $m=5$ moments. The difference between the full order reference mean and the respective ROM prior mean is plotted in gray lines. Notice that for $m=15$ there is no visible deviation from the full order mean, even though a far smaller system of equations was solved. In the $2\sigma$ bands, for $m=15$ only a small error is visible compared to the $m=5$ case.}\label{fig:fig3}
\end{figure}

    In Figure \ref{fig:fig2}, the 1D problem setup is exemplary shown for $\omega/2\pi = 460$Hz. We generate synthetic data, by adding noisy perturbations to a full order reference $\bm{\mu}_{\bm{u};\mathrm{ref}}(\omega)$ with $N_{\mathrm{ref}} = 1000$ DOFs. The data are collected on $n_y=11$ equally spaced sensor locations throughout the domain. Per sensor, $n_o=20$ observations are taken. The FEM prior $\bm{u}$ is the solution of an FE simulation, obtained with the same setting as $\bm{\mu}_{\bm{u};\mathrm{ref}}(\omega)$, with $N=101$ DOFs and QMC sampling. The model error, hence, is only due to discretization. The ROM prior $\tilde{\bm{u}}$, which is shown here for a moment matching ROM with $m=5$ moments matched around $\bar{\omega}/2\pi = 100$Hz, neither follows the data nor the FEM prior. Hence, there clearly is a significant ROM approximation error. This error is supposed to decrease with more matched moments $m$. This is exemplary shown in Figure \ref{fig:fig3}. Notice how a change from $m=5$ to $m=15$ leads to a ROM prior with almost no difference to the full order reference. In the next section, the convergence of the ROM prior towards the FEM prior with increasing $m$ is analysed in more detail.

\subsubsection{ROM convergence}
As the statFEM error estimates were analyzed in \cite{KARVONEN2022} and estimates of the FEM error for the Helmholtz equation have been discussed extensively in the literature \cite{IHLENBURG1995}, we put the main emphasis here on the convergence of the ROM prior approximation and the statROM update.
All errors are specified as the expected value of the wave number dependent $H_k^1$ energy norm \cite{Spence2022}
\begin{equation}
    \|\mu_{u;\mathrm{ref}}- {\mu}_{\tilde{{u}}} \|_{H_k^1}^2 = k^{-2} \| \nabla (\mu_{u;\mathrm{ref}}-{\mu}_{\tilde{{u}}}) \|^2_{L^2} + \| \mu_{u;\mathrm{ref}}-{\mu}_{\tilde{{u}}} \|^2_{L^2}, 
\end{equation}
where discrete quantities are interpolated with the underlying FE basis functions. The lower bound of the error always depends on the wave number, i.e. the physical setting of the problem as shown in \cite{IHLENBURG1995}.

To compute the ROM approximation error, the approximated FE solution serves as a reference.

\begin{figure}[ht]
\begin{center}
\includegraphics[width=0.97\textwidth]{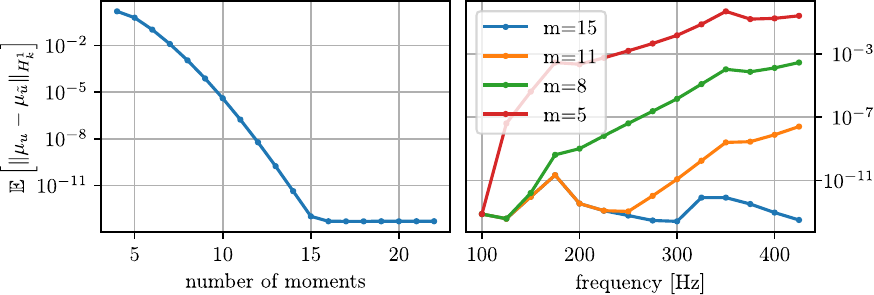}
\end{center}
\caption{Left: The approximation error of the ROM mean quickly converges to a constant quantity at $m=15$ matched moments. Therefore, a ROM with $m\geq16$ can be considered to provide an exact representation of the full order model. For the computation of the error norm, the ROM solution $\tilde{\bm{u}}(\omega)$ and the full order solution $\bm{u}(\omega)$ are evaluated at $\omega/2\pi = 460\mathrm{Hz}$ while the expansion frequency is $\omega/2\pi = 100\mathrm{Hz}$.\\
Right: For higher numbers of matched moments $m$ around $\bar{\omega}/2\pi=100$Hz, the ROM yields smaller errors throughout the frequency range. With $m=15$, the error is similarly low throughout the frequency range.}\label{fig:fig4}
\end{figure}

Figure \ref{fig:fig4} shows the convergence of the moment matching ROM. In the left plot, the expansion is centered around a single expansion frequency $\bar{\omega}/2\pi = 100\mathrm{Hz}$. The ROM solution $\tilde{\bm{u}}(\omega)$ and the full order solution $\bm{u}(\omega)$ are evaluated at $\omega/2\pi = 460\mathrm{Hz}$. One may observe that the ROM converges, in this case, to machine precision at $m = 15$ moments chosen for the expansion.


Choosing a fixed number of moments in the expansion and calculating the error norm for a set of frequencies yields  the result plotted on the right side. As the expansion is centered around $\bar{\omega}/2\pi = 100\mathrm{Hz}$, the error at that frequency is very low, regardless of how many moments are used in the expansion. Clearly, the more moments are used, the lower the error also for frequencies further away from $\bar{\omega}$.

Next, the statROM update is performed and the convergence of the predictive observation density towards the reference solution is analyzed.

\subsubsection{statROM Convergence}
\begin{figure}[ht]
\begin{center}
\includegraphics[width=0.97\textwidth]{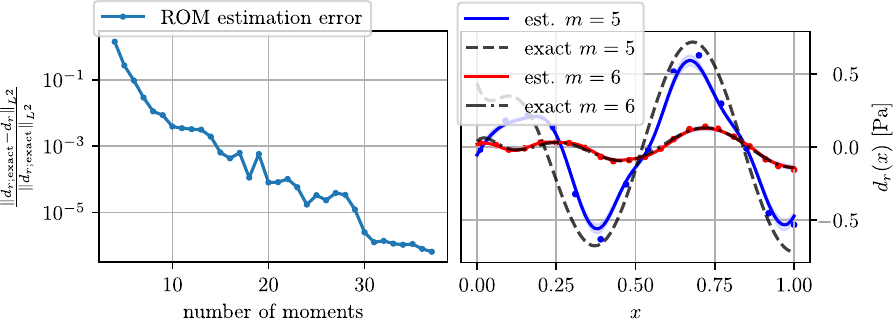}
\end{center}
\caption{Left: Convergence of the ROM error estimator at the training points with increasing number of matched moments. A relative error of $0.1\%$ is reached at about $15$ matched moments.\\
Right: For the example from before with $\bar{\omega}/2\pi=100$Hz, $m=5$, the ROM error is approximated with an adjoint error indicator which is evaluated at $14$ training points. A GP is trained to give an approximation of the ROM error in the whole field. Notice that while the rough shape of the error is well approximated, there is still a significant difference to the exact ROM error.\\
Increasing the number of test points to $22$ and choosing $m=6$ instead of $m=5$ matched moments leads to a better approximation of the ROM error. The amplitude of the error naturally decreases because of the larger ROM basis.}\label{fig:fig5}
\end{figure}

Throughout this section, the performance of the proposed method is compared to a reference, where the standard (statFEM) posterior update is performed on a ROM prior instead of a FEM prior. The prior is conditioned on noisy data drawn from a finely discretized ($1000$ elements) FEM solution of the problem. For the statROM posterior update, the ROM error estimate is used to construct the predictive obervation density, which corrects the biased posterior with the estimated error. In the next paragraphs, the error between the ground truth and the results of the different approaches is evaluated for different ROM fidelities to show the convergence properties.
%
%
\paragraph{Correction with adjoint error estimate}
The adjoint error estimator is evaluated at a set of training points throughout the domain. For each of the points, a ROM projection matrix has been constructed. Figure \ref{fig:fig5} shows the convergence of the ROM error estimate with $12$ training points. One may observe as the number of matched moments grows, not only the ROM itself but also the corresponding error estimator becomes significantly more
accurate. 
This is an interesting fact which sets the bounds in which the proposed method is applicable: For very poor quality ROMs, e.g. as shown in Figure \ref{fig:fig2}, the error estimator will also be of rather low quality and no improvement over the standard method will be obtained. On the other hand, if the ROM has very high accuracy, the ROM error estimator is highly accurate, but the error is so small that it doesn't need to be quantified. The method is most relevant, as will be shown in the next chapters, for ROMs of medium quality. There, the estimation of the ROM error is feasible with a good accuarcy and the errors are still large enough so that an error correction improves the posterior update. 

Figure \ref{fig:fig5}, right side, demonstrates the effect of using more training points for the error GP as well as a higher fidelity ROM. The ROM error decreases and the quality of the estimator increases. The convergence of the estimator at the test points (as opposed to the training points which have been covered in Fig. \ref{fig:fig5}, left side) depends on the chosen GP configuration: Besides the number of training points, also other factors such as the choice of kernel function and hyperparameters have an effect on the absolute error and convergence. The reader is referred to \cite{rasmussen2006} for details on how a GP regression can be optimized. In the following chapters, a Matérn $5/2$ kernel function and frequency dependent hyperparameters are used, i.e.
the length scale
\begin{equation}
    l(\omega) = \frac{\lambda}{4}=\frac{2\pi c}{\omega}
\end{equation}
and the scaling
\begin{equation}
    \sigma(\omega) = 2\;\mathrm{max}(|\bm{d}_r(\omega)|).
\end{equation}



\begin{figure}[ht]
\begin{center}
\includegraphics[]{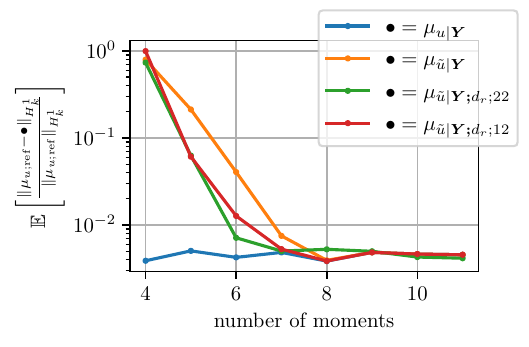}
\end{center}
\caption{The expected value of the error for different sizes of the ROM basis, i.e. different values of $m$. The standard (statFEM) approach applied to the ROM prior, i.e. $\mu_{\tilde{u}|\bm{Y}}$ in orange, yields a slower convergence than the statROM approach with the simplest adjoint error estimator, i.e. $\mu_{\tilde{u}|\bm{Y};d_r;12}$ in red. Notice that increasing the fidelity of the error estimator yields even faster convergence, i.e. $\mu_{\tilde{u}|\bm{Y};d_r;22}$ in green.}\label{fig:fig6}
\end{figure}

In Figure \ref{fig:fig6} the convergence of different ROM predictive observation densities is displayed as a function of the chosen basis size. The baseline error is determined by the full order statFEM predictive density, which still has a certain error compared to the ground truth. This error depends on the remaining discretization error (data are generated on a very fine grid), the finite number of observations per sensor, the defined sensor noise, the choice of norm and also the small diagonal quantity added to the prior covariance matrices for numerical stability. Plotted in bold lines are the errors of the predictive densities constructed with the proposed adjoint ROM error estimator. For comparison, the convergence of the standard approach, i.e without any ROM error estimation is shown, too. Notice that a simple version of the adjoint estimator, i.e. with $12$ training points, already delivers faster convergence than statFEM with a ROM prior only, i.e. without the error estimator.

\begin{figure}[t]
\begin{center}
\includegraphics[]{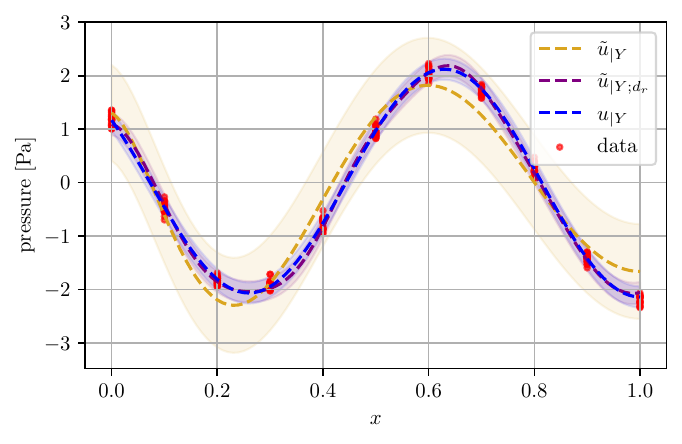}
\end{center}
\caption{A comparison of the resulting posterior solutions from the ROM prior $\tilde{u}_{|Y}$ in yellow, the ROM prior with the error estimator  $\tilde{u}_{|Y;d_r}$ in purple (the proposed statROM approach) and the full order reference prior $u_{|Y}$ in blue. Notice that for the standard approach, i.e.  $\tilde{u}_{|Y}$, the model error is overestimated for low fidelity ROMs because the ROM error is considered model error as well. This results in a large posterior standard deviaton. To illustrate the principles, a bad ROM representation with basis size $m=5$ has been chosen on purpose. One may observe that the estimated ROM error is used as a correction bias for the data (red dots). The data corrected by the error estimator lie much closer to the ROM prior. Hence, the resulting model error for the proposed approach is much smaller than for the standard one, in which the ROM prior is conditioned on uncorrected data. Clearly, using the estimated ROM error to correct the posterior yields a result close to the ground truth even though the ROM prior is used.}\label{fig:fig7}
\end{figure}

The corresponding predictive states for $m=5$ are visible in Figure \ref{fig:fig7}. The ROM error estimate was computed with $12$ training points evenly spaced throughout the domain. In the result for the standard approach, i.e. the yellow curve $\tilde{\bm{u}}_{Y}$, the ROM prior is scaled with $\rho$ to account for the data as well as possible. The estimated model error is quite large as the ROM approximation error is part of it. Hence, the resulting variance is also relatively large compared to the prior and the full order predictive posterior. The statROM approach with the adjoint ROM error estimator, i.e. the purple line $\tilde{\bm{u}}_{Y;d_r}$, results in a smaller model error  and hence, a much smaller predictive posterior variance with a mean much closer to the ground truth.

One may observe that the proposed statROM approach yields more accurate results.
The key insight of the results in Figure \ref{fig:fig7} is the data correction induced by the statROM update. Clearly, the yellow curve (statFEM) is the ROM prior scaled by $\rho$ so that it passes through the measurement data. Because of the ROM-corrected data model, here the resulting marginal likelihood \eqref{eqn:logLikROM}, introduces a bias as large as the ROM error at the data points: the hyperparameters are estimated in order to comply with the \emph{corrected} data, i.e. data close to the ROM prior. Therefore, the estimated model error GP ($\bm{d}$) of $\tilde{\bm{u}}_{Y_{d_r}}$ exhibits a very small $\sigma_d$: The ROM-prior lies close to the corrected data. The variance for the corrected posterior, as visible in Fig. \ref{fig:fig7}, is rather small.

On the other hand, the estimated model error GP $\bm{d}$ of $\tilde{\bm{u}}_{Y}$, i.e. for the uncorrected posterior, exhibits a rather large $\sigma_d$: The (ROM)-prior doesn't lie on the actual measured data. This reflects a large posterior variance as visible in Fig. \ref{fig:fig7}.


Next, a more complex 2D acoustic scattering example is used to test and compare the methods.
\subsection{2D acoustic scattering example}
The setting for the 2D acoustic scattering example differs from the 1D example in the choice of boundary conditions and discretization. We define an unstructured mesh on a square domain. In the middle, a circular scatterer is placed. All outer boundaries are defined to be absorbing. The boundary of the scatterer is defined to be homogeneous Dirichlet. As source term, an incident wave over the whole domain is defined. Contrary to the 1D example, we introduce additional model error by altering the source term defined to generate the ground truth.


\subsubsection{Forward Problem}
The mean of the stochastic source term, i.e. the incident wave, is defined as 
\begin{equation}
    \bar{f}(\omega,\bm{x}) = \mathrm{exp}(\mathrm{i} k x) = \mathrm{cos}(kx) + \mathrm{i}\;\mathrm{sin}(kx).
\end{equation}
The mean of the incident wave is constant along the y axis.
The covariance in both real and imaginary part is modelled with a Mat\'ern kernel \eqref{eqn:MaternKernel} with hyperparameters $\nu = 1.5$, $\sigma_f = 0.8$ and $l_f = 0.6$, hence a GP is defined for both real and imaginary parts of the forcing. Therefore, we have
\begin{align}
    f_{GP_\mathrm{re}}&\sim\mathcal{GP}(0,k_f(\bm{x}))\\
    f_{GP_\mathrm{im}}&\sim\mathcal{GP}(0,k_f(\bm{x}))
\end{align}
and thus
\begin{equation}
    f = \bar{f} + (f_{GP_\mathrm{re}} + \mathrm{i}f_{GP_\mathrm{im}}) = \bar{f} + f_{GP}.
\end{equation}
The stochastic forcing and subsequent sampling lead to an incident wave that is not constant along the y axis any more, compared to the mean.

\begin{figure}[ht]
\centering
\includegraphics[width=0.8\textwidth]{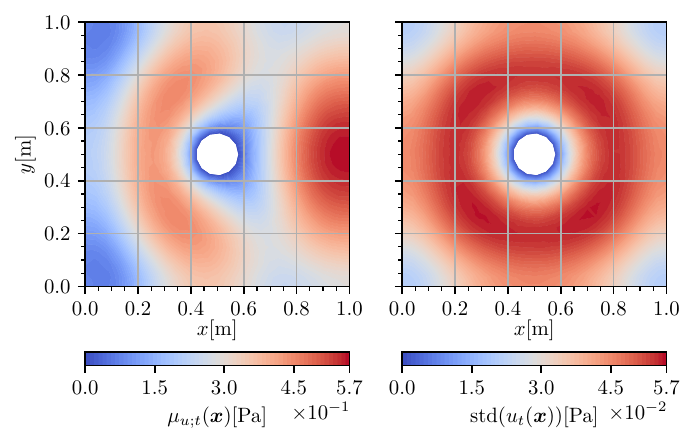}
\caption{Mean and variance of the physical field of the FEM prior. Because of the homogeneous Dirichlet condition at the scatterer, both mean and variance are zero there. On all other boundaries, absorption conditions mostly block reflection.}
\label{fig:fig8}
\end{figure}

All following plots display the modulus, i.e. the total physical field, of the complex valued solution field quantity. I.e. for the FEM prior $\bm{u}_t(\bm{x}) = |\bm{u}(\bm{x})|$.
Solving the stochastic FEM problem with a QMC approach yields the results visible in Figure \ref{fig:fig8}. 

One can observe that both the pressure field and its variance are zero at the Dirichlet boundary of the circular scatterer. The prior computation is the step in the workflow where using the ROM is crucial: The solver time per sample is significantly reduced compared to the full order model. For the 2D example, we measured an average of $0.001938\text{s}$ per sample for the full order model and an average of $0.0006217\text{s}$ per sample for the ROM. This includes reduction of the matrices, solving the reduced system and projecting back onto the full order space. FE system assembly has to be done for both the full order model and the ROM and was excluded from both measurements. Per training point in the estimator, one solution of the reduced order system is performed. Therefore, one has to assess whether the statROM approach is still faster than the full order approach when many error estimator training points are in use. In the case of the 2D example with $256$ parameter samples, the total prior runtime for the full order approach amounts to $6.4382\text{s}$ and for the statROM approach $5.8377\text{s}$, including the error estimator with $200$ training points ($0.29373\text{s}$) and FE assembly. Hence, in this example, $256$ FEM solves are compared to $456$ ROM solves. Note that the computation of the error estimator could be accelerated further by e.g. using sparse GP approximation \cite{sparseGP} and the global Arnoldi method, see Remark~\ref{rmk:AORGA}.

\begin{figure}
\centering
\includegraphics[width=1.0\textwidth]{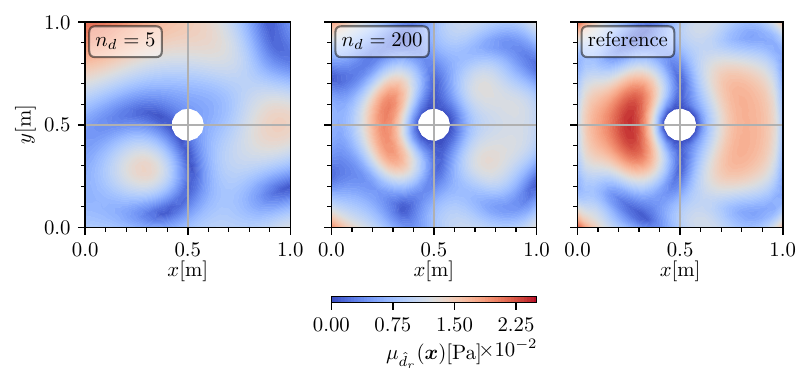}
\caption{The estimate of the ROM error improves with an increasing number of training points. The quality of the estimation, i.e. the difference of the converged estimator to the reference, depends on the quality of the ROM, cf. Fig. \ref{fig:fig5}. In this particular case with $m=12$, an approximation error in the estimator is visible. Increasing $m$ would lead to a better approximation but is linked to higher computational cost.}
\label{fig:fig9}
\end{figure}

For the ROM prior, a moment matching approach with $m=12$ is chosen. The corresponding error to the FEM full order solution is visible in Figure \ref{fig:fig9}. Using the adjoint error indicator, the ROM error is estimated at the training points which are spread evenly throughout the domain. For that, the full order model does not have to be evaluated in the online phase but only the corresponding adjoint ROM, which is cheaper in comparison. 

The resulting mean of the ROM error estimate GP is visible in Figure \ref{fig:fig9} for different sizes of training samples: Clearly, a larger number of samples yields a more accurate estimation. For the following studies, we choose an estimator with $n_d=200$ training points. As the Dirichlet condition at the scatterer is known also in the ROM, the error estimator GP is trained with $\bm{d}_r = 0$ at the corresponding nodes without computing the error estimate at these locations.

\begin{figure}[t!]
\centering
\includegraphics[width=0.8\textwidth]{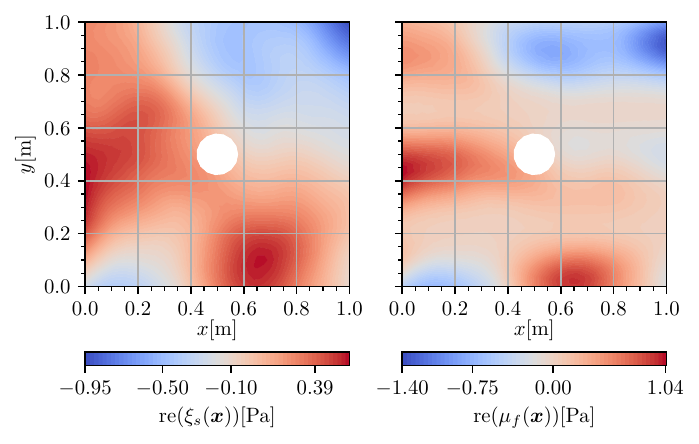}
\caption{Data are generated on a more finely discretized mesh ($1131$ degrees of freedom vs. $854$ for the prior). A parameter sample (left) serves as the basis and is superimposed with a sine wave in orthogonal direction to the incoming wave in the prior forcing (right)}.
\label{fig:fig10}
\end{figure}

In this example, the ground truth is not only a more finely discretized version of the FEM prior but involves a different modelling choice as well. In particular, the ground truth is found by solving the problem defined with the source term
\begin{equation}
   f =  \bar{f} + \mathrm{re}(\xi_s)  (0.8\mathrm{cos}(4.5\pi y) + 1) +\mathrm{i} \;\mathrm{im}(\xi_s)  (0.8\mathrm{sin}(4.5\pi y) + 1) ,
\end{equation}
where $\xi_s(x)$ is a sample of the original stochastic part of the source term $s$. In this way, the sample is superimposed with a sine in the orthogonal direction to the original incoming wave, resulting in an altered scattering behavior. Figure \ref{fig:fig10} shows the real part of the described process: The original sample and the altered one which serves as ground truth.

\begin{figure}
\centering
\includegraphics[width=0.95\textwidth]{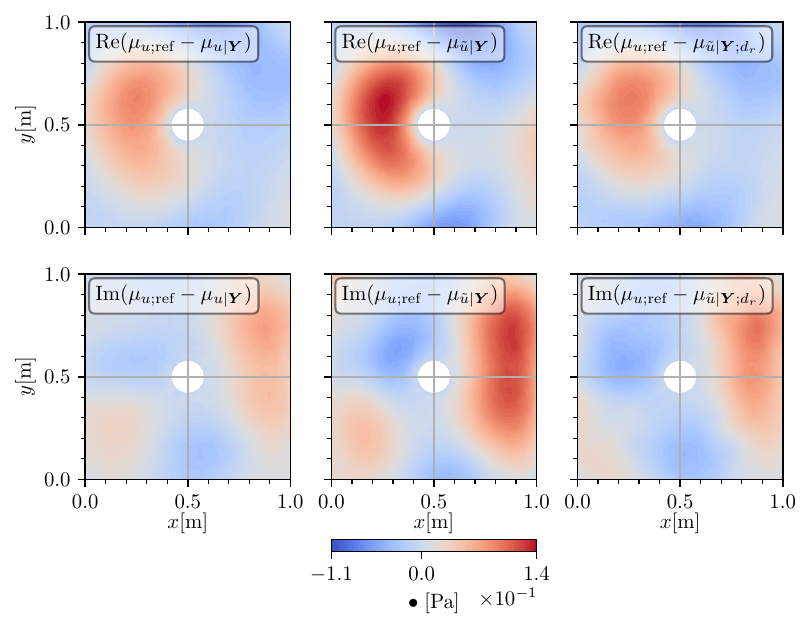}
\caption{The difference of the ground truth (reference) mean $\mu_{u\text{;ref}}$ to the posterior mean of the full order method (left column), the standard statFEM approach applied to a ROM prior (middle column) and the proposed statROM method (right column). We observe that the posterior mean of the proposed reduced order method ($\mu_{\tilde{u} |{\bm{Y};d_r}}$) yields results closer to the full order statFEM method ($\mu_{{u} |\bm{Y}}$) compared to the standard statFEM on a ROM prior ($\mu_{\tilde{u} |{\bm{Y}}}$). This holds true both for the real (top row) and imaginary (bottom row) parts of the solution.}
\label{fig:fig11}
\end{figure}

In Table \ref{tab1}, the posterior error is listed for three different sensor configurations: low resolution,
moderate resolution, and high resolution. The low resolution configuration is defined with $5$ sensors and $20$ observations per sensor, the moderate resolution one with  $30$ sensors and $50$ observations and the high resolution configuration with $n_s = 80$ and $n_o=200$. 
\begin{table}[t!]
\caption{$m=12$, relative posterior error norm for the different methods for different sensor data configurations. The respective posterior mean is compared to the ground truth. The proposed method with an error estimator yields smaller errors compared to the standard method with the ROM prior, if the size of the data set is small. Increasing the number of sensors and observations decreases the weight of the prior and hence a smaller error is achieved for all methods. The difference in the error between the methods is smaller.}\label{tab1}
\begin{tabular*}{\textwidth}{@{\extracolsep\fill}lccccccc}
\toprule%
& \multicolumn{2}{@{}c@{}}{$400 \mathrm{Hz}$, $5$ sens, $20$ obs } & \multicolumn{2}{@{}c@{}}{$400 \mathrm{Hz}$, $30$ sens, $50$ obs} & \multicolumn{2}{@{}c@{}}{$400 \mathrm{Hz}$, $80$ sens, $200$ obs} \\\cmidrule{2-3}\cmidrule{4-5}\cmidrule{6-7}%
Method &  err re & err im &   err re & err im &  err re & err im\\
\midrule
FOM  & $2.89\mathrm{e}{-2}$ & $1.77\mathrm{e}{-2}$            & $1.03\mathrm{e}{-2}$ & $8.75\mathrm{e}{-3}$   & $1.03\mathrm{e}{-2}$ & $7.70\mathrm{e}{-3}$\\
w/o est  & $5.76\mathrm{e}{-2}$& $4.34\mathrm{e}{-2}$         & $1.14\mathrm{e}{-2}$ & $1.06\mathrm{e}{-2}$   & $1.10\mathrm{e}{-2}$ & $7.98\mathrm{e}{-3}$\\
With est  & $3.19\mathrm{e}{-2}$ & $2.63\mathrm{e}{-2}$       & $1.04\mathrm{e}{-2}$ & $1.05\mathrm{e}{-2}$   & $1.07\mathrm{e}{-2}$ & $8.02\mathrm{e}{-3}$\\
\botrule
         &  $\sigma_d$ re & $\sigma_d$ im&  $\sigma_d$ re & $\sigma_d$ im &$\sigma_d$ re & $\sigma_d$ im\\
\midrule
FOM        & $2.17\mathrm{e}{-2}$& $3.44\mathrm{e}{-2}$      & $1.93\mathrm{e}{-3}$& $7.15\mathrm{e}{-3}$   & $2.96\mathrm{e}{-3}$& $9.72\mathrm{e}{-3}$\\
w/o est    & $1.96\mathrm{e}{-2}$& $4.52\mathrm{e}{-2}$      & $1.60\mathrm{e}{-2}$& $1.25\mathrm{e}{-2}$   & $1.78\mathrm{e}{-2}$& $7.99\mathrm{e}{-3}$\\
With est   & $2.29\mathrm{e}{-2}$ & $3.51\mathrm{e}{-2}$     & $1.29\mathrm{e}{-2}$ & $6.99\mathrm{e}{-3}$  & $3.92\mathrm{e}{-3}$ & $5.87\mathrm{e}{-3}$\\
\botrule
\end{tabular*}
\end{table}
Note that in the low resolution example the remaining error is lowest for the full order problem and highest for the standard statFEM on ROM prior approach, i.e. without the error estimator. The proposed approach with the error estimator yields better results as the standard one. When the number of sensors and observations is increased, the data gain more weight in the Bayesian update compared to the prior and the errors are smaller and more similar throughout the different methods. For the estimated model error standard deviation, the results are very similar throughout the methods. For the low resolution configuration, the differences of the respective posterior state to the ground truth, i.e. reference solution, are plotted in Figure \ref{fig:fig11}. Notice that the difference is the largest for the standard statFEM performed on the ROM prior. The proposed approach yields a difference closer to the full order approach. 

\begin{figure}
\centering
\includegraphics[]{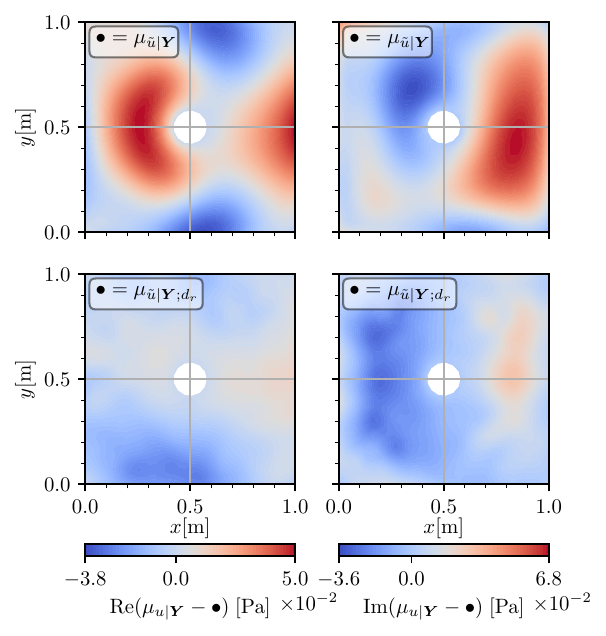}
\caption{The posterior mean difference in the low resolution sensor configuration of the full order solution and the standard statFEM approach (top row) compared to the proposed statROM approach (bottom row). We observe that the posterior mean of the proposed reduced order method ($\mu_{\tilde{u} |{\bm{Y};d_r}}$) yields results much closer to the full order statFEM method ($\mu_{{u} |\bm{Y}}$) than the standard statFEM ($\mu_{\tilde{u} |{\bm{Y}}}$) does. This holds true both for the real and imaginary parts of the solution.}
\label{fig:fig12}
\end{figure}

In Table \ref{tab2}, the posterior error is listed with the full order posterior as reference. As before, we observe that the proposed approach yields a lower error compared to the standard approach. The difference plots for the low resolution configuration in Figure \ref{fig:fig12} support this observation. 
\begin{table}[t]
\caption{$m=12$, relative posterior error norm for the different methods for different sensor data configurations. The respective posterior mean is compared to the full order counterpart. The proposed method with an error estimator yields smaller errors compared to the statFEM method with a ROM prior, when the data size is rather small. By increasing the number of sensors and observations, the weight of the prior is decreased and hence, a smaller error is achieved for all methods. Also the difference in the error between the methods is smaller.}\label{tab2}
\begin{tabular*}{\textwidth}{@{\extracolsep\fill}lccccccc}
\toprule%
& \multicolumn{2}{@{}c@{}}{$400 \mathrm{Hz}$, $5$ sens, $20$ obs } & \multicolumn{2}{@{}c@{}}{$400 \mathrm{Hz}$, $30$ sens, $50$ obs} & \multicolumn{2}{@{}c@{}}{$400 \mathrm{Hz}$, $80$ sens, $200$ obs} \\\cmidrule{2-3}\cmidrule{4-5}\cmidrule{6-7}%
Method &  err re & err im &   err re & err im &  err re & err im\\
\midrule
w/o est  & $1.23\mathrm{e}{-2}$& $1.03\mathrm{e}{-2}$         & $1.18\mathrm{e}{-3}$ & $2.11\mathrm{e}{-3}$   & $5.14\mathrm{e}{-4}$ & $5.49\mathrm{e}{-4}$\\
With est  & $1.73\mathrm{e}{-3}$ & $3.27\mathrm{e}{-3}$       & $5.37\mathrm{e}{-4}$ & $1.22\mathrm{e}{-3}$   & $3.87\mathrm{e}{-4}$ & $6.88\mathrm{e}{-4}$\\
\botrule
\end{tabular*}
\end{table}

\begin{figure}
\centering
         \includegraphics[width=0.85\textwidth]{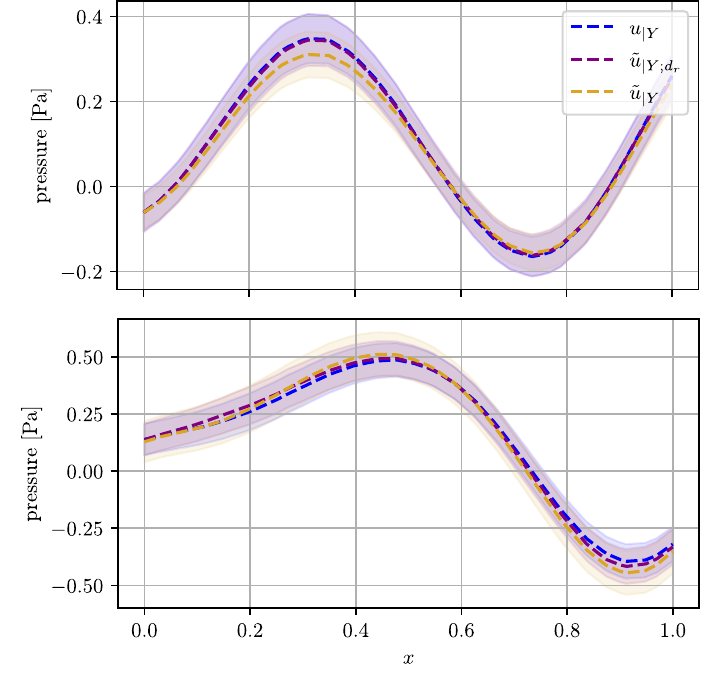} 
         \caption{A slice in x-direction through the posterior states at $y=0.75$. Top: Real part, Bottom: Imaginary part. The full order reference is plotted in blue, the standard statFEM method applied to a ROM prior is plotted in yellow, and the proposed approach is plotted in purple. We observe that the mean of the proposed statROM approach comes closer to the full order posterior than the standard approach does. Notice that the standard deviation is similar for all methods.}
         \label{fig:fig13}
\end{figure}

In Figure \ref{fig:fig13}, a slice through the 2D posterior states is plotted so that the difference between the proposed method, the standard method and the full order reference can be easily observed. Although the errors are small compared to the 1D example, notice that the mean of the proposed statROM method approaches the mean of the full order method, whereas the error for the standard approach is higher. The variance is similar throughout the different methods.



Choosing another setting, i.e. a ROM with more moments matched ($m=20$) and a frequency of $300$Hz, i.e. closer to the expansion frequency of $250$Hz, the ROM already provides a good representation of the underlying full order model and the different methods yield very similar errors. This means that, although the ROM error is estimated accurately, there is no advantage of using the estimator in the inference because the error of the ROM prior is already very small. We omit a table for this example.

In Table \ref{tab3}, the posterior error is listed with the full order posterior as reference for the $300 \mathrm{Hz}$, $m=20$ setting where we also observe no large differences between the methods.
\begin{table}[ht]
\caption{$m=20$, relative posterior error norm for the different methods for different sensor data configurations. The respective posterior mean is compared to the full order counterpart. The proposed method with an error estimator yields smaller errors compared to the standard statFEM method with the ROM prior, in case of a small data set. Increasing the number of sensors and observations decreases the weight of the prior and hence a slightly smaller error is achieved for all methods.}\label{tab3}
\begin{tabular*}{\textwidth}{@{\extracolsep\fill}lccccccc}
\toprule%
& \multicolumn{2}{@{}c@{}}{$300 \mathrm{Hz}$, $5$ sens, $20$ obs } & \multicolumn{2}{@{}c@{}}{$300 \mathrm{Hz}$, $30$ sens, $50$ obs} & \multicolumn{2}{@{}c@{}}{$300 \mathrm{Hz}$, $80$ sens, $200$ obs} \\\cmidrule{2-3}\cmidrule{4-5}\cmidrule{6-7}%
Method &  err re & err im &   err re & err im &  err re & err im\\
\midrule
w/o est  & $3.02\mathrm{e}{-6}$& $1.43\mathrm{e}{-5}$         & $2.53\mathrm{e}{-6}$ & $1.55\mathrm{e}{-5}$   & $1.60\mathrm{e}{-6}$ & $5.79\mathrm{e}{-6}$\\
With est  & $1.42\mathrm{e}{-6}$ & $2.33\mathrm{e}{-6}$       & $2.84\mathrm{e}{-6}$ & $3.75\mathrm{e}{-6}$   & $8.90\mathrm{e}{-7}$ & $2.33\mathrm{e}{-6}$\\
\botrule
\end{tabular*}
\end{table}

We want to summarize the findings from the 2D example and point out the most important building blocks of the method as well as the approximations that had to be made:

\begin{enumerate}
    \item QMC prior construction
        \begin{itemize}
            \item The QMC method approximates the solution statistics of the given problem. Approximation errors occur because of a limited sample size and the Gaussian assumption.
            \item For all QMC samples, a ROM basis was constructed which serves as a fast-to-solve approximation to the full order counterpart.            
        \end{itemize}
    \item ROM error estimation
        \begin{itemize}
            \item The ROM approximation error was estimated using the adjoint approach. An adjoint ROM basis was constructed in the offline phase. 
            \item Errors occur because of a limited adjoint ROM basis size and a limited amount of training points for the GP regression.            
        \end{itemize}
    \item Data Assimilation
        \begin{itemize}
            \item Here we compared three different settings. 1. The full order statFEM, 2. statFEM using the ROM prior without an error estimator and 3. The proposed statROM approach with an error estimator. 
            \item All settings suffer from noisy data. Also, with limited data and a mismatch between prior and data, there is still an error between the posterior and the ground truth. 
            \item In the chosen settings, the ROM prior exhibits a stronger mismatch to the data, which is why the posterior error is higher than for the full order method.
            \item The key finding of this paper is that the proposed statROM method is able to mitigate that higher error to some extend. This makes it possible to solve the QMC ensemble more quickly using the ROM, while maintaining a high accuracy in the posterior. 
        \end{itemize}
\end{enumerate}


\section{Conclusion}\label{sec13}

A ROM adaptation of statFEM has been developed. The Bayesian prior is not constructed using FEM but with a ROM based on a moment matching approach. This is necessary if the frequency dependent full order model is too expensive to be solved at many frequency steps. Compared to the standard statFEM approach applied to a ROM prior, statROM performs better because the ROM error is estimated beforehand. This was shown by computing the error and convergence properties of the standard and the proposed approach compared to a full order reference solution. The new method enables solving the prior in frequency sweeps much faster as it is the case for a full order model while estimating the ROM error.

The estimation of ROM error is achieved using an adjoint estimator, which notably circumvents the need for inverting the full order system matrix. This estimated error is then incorporated into the statistical data model formulation. Introducing this estimated error into the data model has been shown to reduce the overall estimated model error and bring the model's mean closer to the ground truth.

For systems with a high dimensional material coefficient, approximated with the Karhunen–Loève expansion, a QMC approach led to an accurate prior formulation.
This methodology has been applied and its convergence properties showcased through a 1D Helmholtz problem. In a more complex 2D scattering problem with significantly more degrees of freedom and a stochastic forcing defined as a random field, it could be shown that the method can be scaled to some extend.

As a next step, efforts are directed towards parameterizing the ROM, which currently can be applied primarily to frequency sweeps. Another important extension to the method is finding a way to keep the memory requirements low, especially for the covariance matrices in larger systems. As soon as parallelization on the frequency grid is applied, several large matrices have to be kept in memory. For instance, the approach by \cite{koh2023}, where sparse precision matrices replace their dense covariance counterpart, could be adapted to work in the setting presented here. Also, a practical application in aerospace acoustics is considered as a feasible applied problem for statROM.
\backmatter

\section*{Declarations}
\begin{itemize}
\item Funding\\
The research leading to these results received funding from the Deutsche Forschungsgemeinschaft (DFG, German Research Foundation) - Projektnummer 255042459 GRK2075/2.
\item Competing interests\\ 
The authors have no financial or proprietary interests in any material discussed in this article.
\item Ethics approval and consent to participate:\\
Not applicable
\item Consent for publication:\\
Not applicable
\item Data availability:\\
All data generated or analysed during this study are included in this published article.
\item Materials availability:\\
Not applicable
\item Code availability:\\
All code was implemented in Python using FEniCSx \cite{BarattaEtal2023,ScroggsEtal2022,BasixJoss,AlnaesEtal2014} as FEM backend. The source code to generate the results in this publication is available at https://github.com/herluc/statROM and on Zenodo https://doi.org/10.5281/zenodo.15746081 \cite{hermann_2024_software}.
The code is able to reproduce both the 1D and the 2D scattering example. It returns information similar to what is presented in Tables \ref{tab1}, \ref{tab2} and \ref{tab3} as well as a selection of plots. The user can set the most important parameters and observe the change in the posterior solution quality. One limitation of the code lies in its ability to generalize to other examples: Right now, only the given Helmholtz PDE can be used on the given geometries, so it would take significant effort to switch to another PDE. Also, the convergence studies are not automated within the code.

\item Author contribution:\\
Conceptualization and Methodology: Lucas Hermann, Ulrich Römer; Formal analysis, investigation, software, validation and visualization: Lucas Hermann; Writing - original draft preparation: Lucas Hermann; Writing - review and editing: Lucas Hermann, Ulrich Römer, Matthias Bollhöfer; Funding acquisition, project administration and resources: Ulrich Römer; Supervision: Ulrich Römer, Matthias Bollhöfer.
\end{itemize}


\bibliography{sn-bibliography}

\end{document}